\PassOptionsToPackage{table}{xcolor}
\documentclass[acmsmall, screen, nonacm]{acmart}
\usepackage{pgfplots}
\pgfplotsset{compat=1.18}
\usepackage[english]{babel}
\usepackage{subcaption}
\usepackage{enumitem}
\usepackage{multirow}
\usepackage{booktabs}
\usepackage{tabularx}
\usepackage{makecell}
\usepackage{xcolor}
\usepackage{algorithm}
\usepackage{algpseudocode}
\usepackage{pifont}
\usepackage[most]{tcolorbox}
\usepackage{xspace}

\newcommand{\heatmaplevel}[1]{%
  \ifnum#1=1 \cellcolor{orange!0}\fi
  \ifnum#1=2 \cellcolor{orange!5}\fi
  \ifnum#1=3 \cellcolor{orange!10}\fi
  \ifnum#1=4 \cellcolor{orange!15}\fi
  \ifnum#1=5 \cellcolor{orange!20}\fi
  \ifnum#1=6 \cellcolor{orange!25}\fi
  \ifnum#1=7 \cellcolor{orange!40}\fi
}

\setcopyright{none}
\newcommand{\RequireH}[1]{%
\Statex \parbox[t]{\linewidth}{\textbf{Require:} #1}%
}

\newcommand{\MODAIC}{\texttt{MODAIC}\xspace}

\begin{document}
\sloppy

\title{\MODAIC: Coordinated Asynchrony for Decentralized Multimodal Learning}

\author{Yanhang Shi}
\affiliation{%
  \department{Department of Electrical and Computer Engineering}
  \institution{Stony Brook University}
  \city{Stony Brook}
  \state{NY}
  \country{USA}}
\email{yanhang.shi@stonybrook.edu}

\author{Xiaoyu Wang}
\affiliation{%
  \department{Department of Electrical and Computer Engineering}
  \institution{New York University}
  \city{Brooklyn}
  \state{NY}
  \country{USA}}
\email{xw2597@nyu.edu}

\author{Houwei Cao}
\affiliation{%
  \department{Department of Computer Science}
  \institution{New York Institute of Technology}
  \city{New York}
  \state{NY}
  \country{USA}}
\email{hcao02@nyit.edu}

\author{Jian Li}
\affiliation{%
  \department{Department of Applied Mathematics and Statistics and Department of Computer Science}
  \institution{Stony Brook University}
  \city{Stony Brook}
  \state{NY}
  \country{USA}}
\email{jian.li.3@stonybrook.edu}

\author{Yong Liu}
\affiliation{%
  \department{Department of Electrical and Computer Engineering}
  \institution{New York University}
  \city{Brooklyn}
  \state{NY}
  \country{USA}}
\email{yongliu@nyu.edu}

\renewcommand{\shortauthors}{Shi et al.}

\begin{abstract}
Multimodal decentralized learning enables collaboration across heterogeneous edge devices without sharing raw data, but differing modality sets create coupled computation and communication bottlenecks. Multimodal agents support cross-modal alignment while training more model components and sharing bandwidth across multiple modality overlays. Under bandwidth limits, these agents can become hot spots: transfers miss aggregation windows, and their updates become stale or underrepresented, degrading worst-agent accuracy. Asynchrony removes global blocking but does not by itself coordinate local cross-modal interaction with timely peer-update exchange.
We present \MODAIC (\underline{MO}dality-\underline{D}ecoupled \underline{A}synchrony with \underline{I}nterleaved \underline{C}oordination), a system that coordinates local modality progress, peer-update delivery, and aggregation under heterogeneous compute and bandwidth. Hierarchical asynchrony lets modalities progress independently while enforcing reciprocal feature exchange; load-aware update fetching overlaps training with communication, limits concurrent serving load, and reuses late arrivals; and consensus-aware peer selection uses lightweight metadata to account for update availability, peer participation, and loss similarity without active probing. Local training proceeds without waiting for remote updates, while model exchange remains peer-to-peer. We implement \MODAIC in a containerized multi-agent framework and evaluate it on controlled local and cross-region cloud testbeds. Across AVE and KU-HAR, \MODAIC improves average accuracy and multimodal worst-agent performance. On AVE at 200~Mbps per agent, it improves multimodal worst-agent accuracy by 11.0 percentage points at 1000 seconds over the strongest baseline. On the cloud testbed, it reduces time-to-accuracy by up to 30\%. Ablations show that the tested load-balancing and delayed-aggregation variants retain a worst-agent accuracy gap, while stress tests demonstrate tolerance to WAN variability and partial tracker unavailability.
\end{abstract}

\ccsdesc[500]{Computing methodologies~Machine learning}
\ccsdesc[500]{Computer systems organization~Peer-to-peer architectures}
\ccsdesc[500]{Networks~Network performance evaluation}

\keywords{decentralized learning, multimodal learning, asynchronous training, modality heterogeneity}

\maketitle

\section{Introduction}\label{sec:intro}

Modern edge applications increasingly rely on data from multiple modalities, such as audio, video, and text, to improve performance and robustness~\cite{ouyang2022cosmo,li2022motion}. At the same time, privacy concerns and data-governance requirements motivate collaborative learning that keeps raw data local~\cite{McMahan2017FedAvg,abdelmoniem2023refl,liu2025practical}. Geographically distributed deployments can additionally benefit from avoiding traffic concentration at a single model-aggregation server. These considerations motivate \emph{multimodal decentralized learning}: agents collaboratively train modality-specific components of a shared task model by exchanging updates directly with peers, even though each agent may observe only a subset of modalities. Agents sharing a modality form a modality-specific exchange overlay, while agents with multiple modalities additionally optimize cross-modal alignment to couple their local representations.

Beyond conventional client-side resource heterogeneity, differences in available modalities introduce a structured source of computation and communication imbalance. Multimodal agents enable cross-modal knowledge transfer, but they also train more model components and participate in multiple modality overlays. The same modality memberships therefore link an agent's role in cross-modal learning to its resource demands. These ``bridge'' agents connect modality-specific learning through local alignment, yet can become slower update producers and communication hot spots under shared resource budgets. Preserving their contribution is important because delays can limit the propagation of cross-modal information as well as the progress of individual agents.

\textbf{What breaks under modality heterogeneity.}
Cooperative multimodal learning introduces both \textit{intra-agent} and \textit{inter-agent} couplings. First, cross-modal alignment and fusion couple modality-specific encoders during training. On multimodal agents, synchronization between modalities can create bursty compute phases and idle gaps, slowing the production of sharable same-modality updates (Challenge~1). Second, multimodal agents must serve and fetch updates for multiple model components through the same bandwidth budget. Concurrent exchanges across overlapping modality groups can create communication hot spots and cause transfers to miss local aggregation windows (Challenge~2). Third, under asynchronous aggregation, each agent must choose which peers to contact and which available updates to incorporate under a limited exchange budget. Contacting more peers can improve information diffusion but increase contention, whereas contacting too few can slow diffusion and concentrate contributions among a small set of agents (Challenge~3). Together, \textit{these effects influence both the age of incorporated updates and whose updates participate in learning, potentially widening accuracy disparities across agents.} Section~\ref{sec:motivation} characterizes these effects through controlled microbenchmarks and end-to-end experiments.

\textbf{Why existing and simpler approaches fall short.}
Prior designs address different parts of this problem. Server-coordinated multimodal training~\cite{ouyang2023harmony,wang2024fedhkd,zheng2023autofed} addresses modality and representation heterogeneity, but its model exchange relies on centralized aggregation and coordination. Asynchronous decentralized schemes~\cite{ad-psgd,bornstein2023swift,asydfl} reduce synchronization overhead and accommodate heterogeneous participants, but do not directly address reciprocal cross-modal interaction within agents together with contention across overlapping modality groups. Peer selection also faces a practical constraint: active probing~\cite{asydfl} adds control messages and can delay scheduling decisions on high-latency links, while random or fixed-topology exchanges do not explicitly account for changing update availability and serving load.

Common systems mechanisms address distinct decisions within this setting. Load balancing reduces transfer contention, delayed aggregation reuses late updates, and peer selection determines which updates to request. Their effectiveness also depends on local modality progress: accelerating one modality changes the availability of its updates, while contention determines whether those updates arrive in time to be incorporated. Network-level scheduling must also coexist with the matched, bidirectional feature exchange needed for local cross-modal learning. These dependencies motivate a runtime that \textit{coordinates local modality progress, peer-update delivery, and aggregation under shared resource constraints}. Our evaluation examines how much of the resulting benefit can be recovered by simpler load-balancing and delayed-aggregation variants.

\textbf{Design requirements.}
The runtime should allow modalities to progress independently while preserving reciprocal cross-modal interaction. It should manage contention and reuse useful late arrivals without blocking local training, and guide peer exchange using lightweight information about update availability and participant state. We therefore ask:
\begin{quote}
\textit{How can we coordinate local cross-modal interaction and peer-update exchange to improve learning under heterogeneous compute and limited bandwidth, without imposing global synchronization?}
\end{quote}

\textbf{Our approach: \MODAIC.}
We answer this question with \MODAIC (\underline{MO}dality-\underline{D}ecoupled \underline{A}synchrony with \underline{I}nterleaved \underline{C}oordination), a wait-free training runtime for multimodal decentralized learning. Here, wait-free means that local training proceeds without waiting for partner features or remote model transfers. \MODAIC follows one principle: \emph{decouple modality progress while coordinating reciprocal local interaction and bandwidth-limited peer-update exchange}. It realizes this principle through three mechanisms. (1) \textit{Hierarchical Asynchronous Collaboration} lets modalities train and aggregate independently, while enforcing matched, contribute-before-consume feature exchange to preserve bidirectional cross-modal interaction (Section~\ref{subsec:hier_async}). (2) \textit{Load-Aware Update Fetching} overlaps communication with computation, limits concurrent serving load through lightweight admission control, and retains late arrivals in per-modality caches for subsequent aggregation (Section~\ref{subsec:selective_peer_design}). (3) \textit{Consensus-Aware Peer Selection} uses lightweight metadata about update availability, sampling history, and loss similarity to guide peer exchange without active probing (Section~\ref{subsec:peer_selection_design}). A centralized metadata tracker supports admission control and peer selection, while all model states are exchanged directly between agents, without centralized aggregation or model-data relaying.

\textbf{Results.}
We implement \MODAIC as a containerized multi-agent training framework and evaluate it on AVE and KU-HAR using controlled local and cross-region cloud testbeds. On AVE at 1000s with a per-agent bandwidth budget of 200Mbps, \MODAIC achieves unimodal and multimodal worst-agent accuracies of 35.16\% and 56.28\%, improving the best corresponding baseline results by 2.99 and 11.01 percentage points, respectively. Relative to AsyDFL, the multimodal average-worst accuracy gap decreases from 7.35 to 3.46 percentage points. On the cross-region cloud testbed, \MODAIC reduces time-to-accuracy by up to 30\%. Additional experiments show that the tested load-balancing and delayed-aggregation variants retain a worst-agent accuracy gap, and examine the contributions of the runtime components. We also quantify tracker overhead and show that performance gains persist under the evaluated WAN-like RTT variation and packet loss, with graceful degradation under tracker delay and partial unavailability.

Our main contributions are summarized as follows:
\begin{itemize}[leftmargin=1.25em,itemsep=2pt,topsep=2pt]
\item We characterize how modality heterogeneity couples local training delays with competition for shared bandwidth across overlapping modality groups, and examine the resulting effects on update exchange and worst-agent performance.
\item We design \MODAIC, a runtime that coordinates asynchronous local cross-modal interaction, load-aware update fetching, and metadata-guided peer selection while preserving reciprocal feature exchange and allowing local training to proceed without waiting for remote updates.
\item We build an end-to-end implementation and demonstrate improvements in learning quality at fixed resource budgets and in time-to-accuracy, while evaluating component contributions and sensitivity to stragglers, WAN dynamics, and tracker availability.
\end{itemize}

\section{Motivation} \label{sec:motivation}
We first show \textit{how modality heterogeneity creates coupled computation and communication bottlenecks in decentralized learning.} The central issue is that multimodal ``bridge'' agents enable cross-modal knowledge transfer while being disproportionately exposed to these bottlenecks.

\textbf{Problem setting.}
We study \emph{multimodal decentralized learning}, where agents hold \emph{heterogeneous modality sets} (e.g., audio, video, or audio+video) and exchange locally updated model parameters in \emph{peer-to-peer} fashion, without centralized model aggregation or model-data relay. For each modality, an exchange overlay is formed for mixing same-modality parameters. Multimodal agents therefore participate in multiple overlays and also perform local \emph{cross-modal interaction/alignment} to couple their representations. This setting links modality-dependent training costs with communication load across overlapping overlays, creating \emph{structured} imbalance in update production and exchange.

\textbf{Key systems intuition.}
Efficient decentralized optimization depends on local updates diffusing through the network \emph{quickly} and \emph{broadly}. Two system-level factors shape this process: (1) \emph{staleness}, the age of an update when it is incorporated~\cite{pmlr-v238-even24a}, and (2) \emph{mixing}, how rapidly information diffuses across agents through successive exchanges~\cite{hua2022efficient}. Under modality heterogeneity, compute and bandwidth imbalances can increase staleness and skew participation, widening accuracy gaps across agents. This is particularly consequential when underrepresented contributors are multimodal agents whose updates carry information learned through cross-modal interaction. We therefore examine \emph{worst-agent performance} alongside average accuracy.

\textbf{Observations and challenges.}
We first characterize two resource effects: (i) delayed update production from modality-dependent computation and cross-modal coupling (Section~\ref{subsec:motivation_compute}), and (ii) load-imbalanced communication from participation in multiple overlays (Section~\ref{subsec:motivation_comm}). Together, these effects create (iii) a dynamic tension among mixing, staleness, and contention when selecting peers under a limited exchange budget (Section~\ref{subsec:motivation_sched}).

\subsection{Computation Heterogeneity}\label{subsec:motivation_compute}

A fundamental obstacle in multimodal decentralized learning is that
agents with more modalities can take longer to produce updated model components, even under the same hardware budget. This affects how frequently other agents can obtain new same-modality contributions from them.

\begin{figure}[t]
    \centering
    \begin{subfigure}[t]{.43\linewidth}
    \centering
    \includegraphics[width=0.7\linewidth]{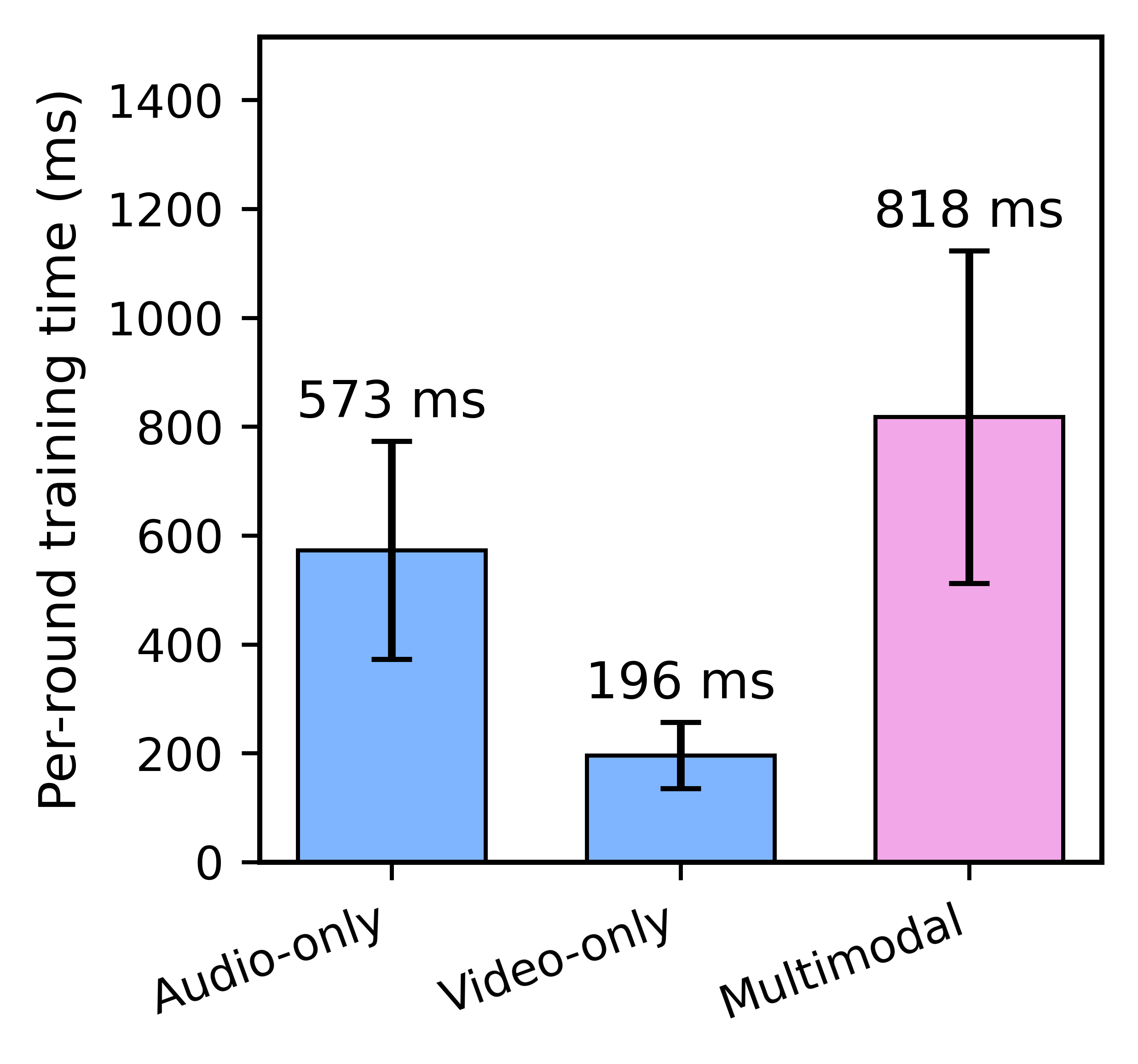}
       \vspace{-0.1in}
   \caption{Per-step training time (same GPUs and neural nets).}
    \label{fig:comp_hetero}
    \end{subfigure}
    \hfill
    \begin{subfigure}[t]{.54\linewidth}
    \centering
    \includegraphics[width=0.7\linewidth]{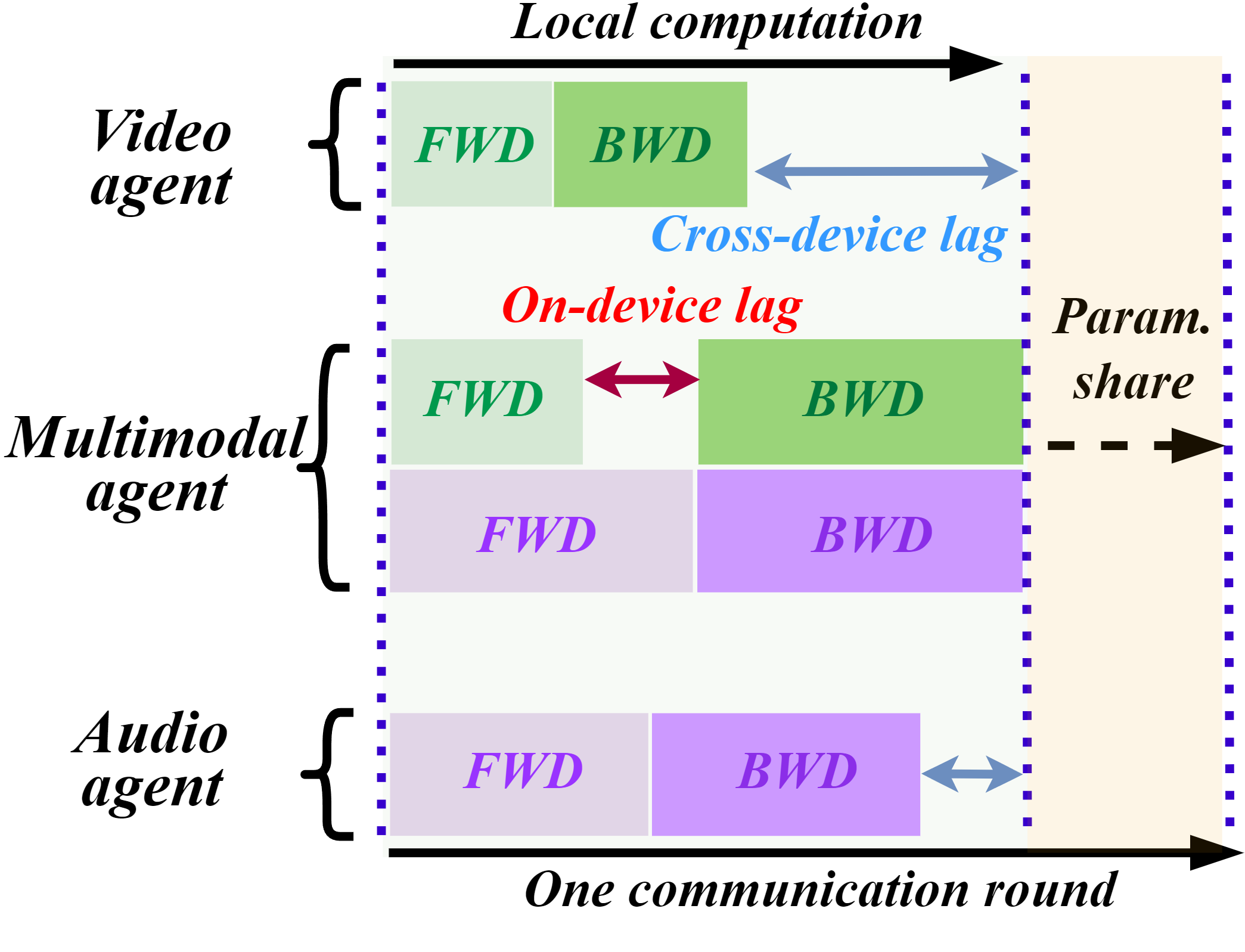}
       \vspace{-0.1in}
    \caption{Synchronized execution timeline (FWD/BWD) highlighting idle gaps.}
    \label{fig:comp_pattern}
    \end{subfigure}
    \vspace{-0.1in}
\caption{Modality-induced computation heterogeneity. (a) Multimodal agents have higher per-step latency than unimodal agents. (b) Cross-modal synchronization creates idle gaps between forward (FWD) and backward (BWD), leading to bursty compute.}
    \label{fig:motivation_hetero}
      \vspace{-0.15in}
\end{figure}

\textbf{Modality-dependent training creates unequal update rates.}
Figure~\ref{fig:motivation_hetero}(a) isolates computation heterogeneity by constructing three groups with different modality sets (audio-only, video-only, and audio+video), using encoders with matched parameter size on identical GPUs. Even in this controlled setting, multimodal agents incur substantially longer per-step training time. As a result, within a same-modality overlay, unimodal agents (e.g., video-only) can receive new video updates from unimodal peers much more frequently than from multimodal peers.
At a fixed wall-clock aggregation cadence, the longer refresh intervals of multimodal agents can leave peers with older available updates, even before transfer delay is considered.
In realistic deployments where modality encoders are not size-matched, the gap can be larger~\cite{distmm24}.

\textbf{Why cross-modal synchronization delays update production.}
Figure~\ref{fig:motivation_hetero}(b) illustrates how synchronized cross-modal interaction contributes to the latency gap. A unimodal step forms a tight loop: forward pass (FWD) followed immediately by backward pass (BWD). In the synchronized multimodal execution shown, backward computation for the joint alignment/fusion loss waits until all required modalities complete their FWD passes~\cite{radford2021learning,jia2021scaling}. This creates modality-level idle gaps: a faster modality finishes its FWD but waits for its partner's features before proceeding with BWD. These dependencies delay the production of shareable same-modality updates. Making peer exchange asynchronous does not, by itself, remove this within-agent synchronization.

\begin{figure}[t]
    \centering
    \includegraphics[width=.7\linewidth]{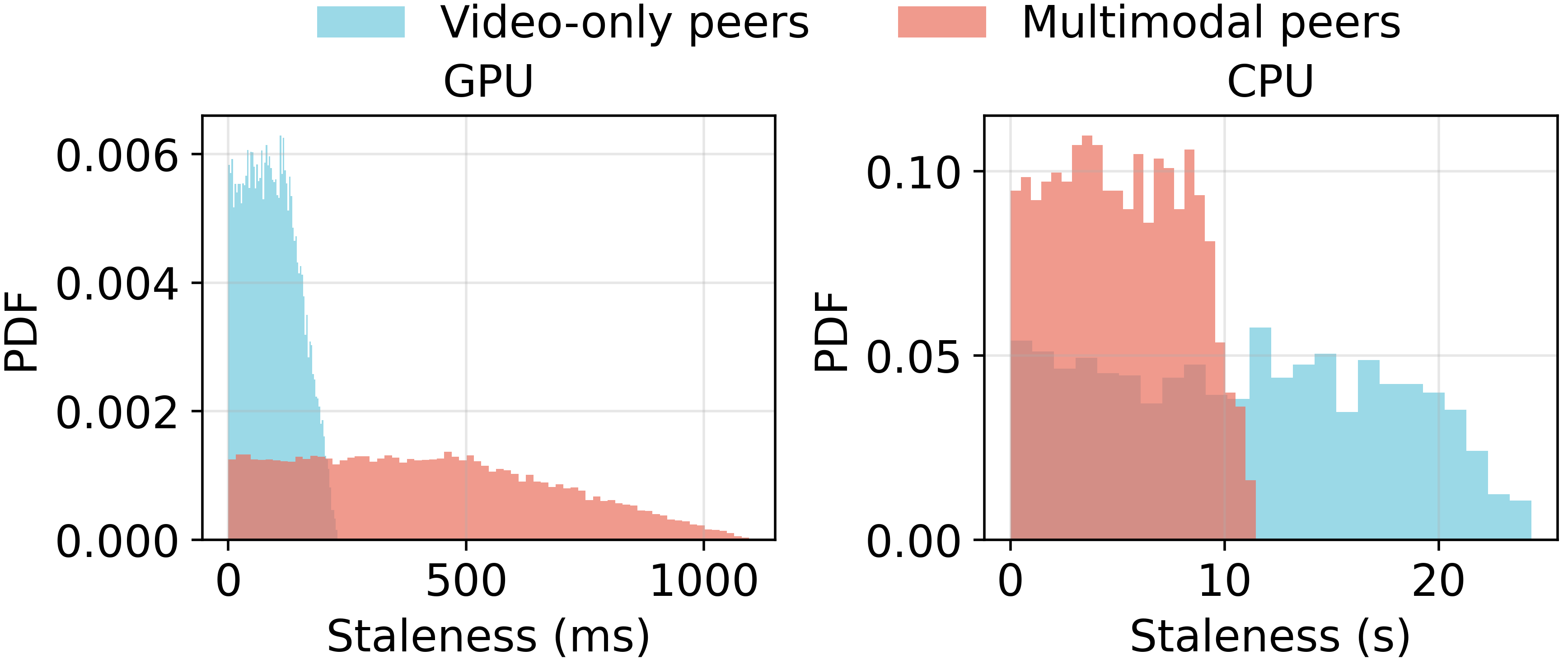}
     \vspace{-0.1in}
    \caption{Update-age distributions at aggregation for video-only agents, comparing updates from video-only and multimodal peers under GPU- and CPU-driven training. Ages are reported in milliseconds and seconds, respectively.}
    \label{fig:staleness_pdf}
     \vspace{-0.15in}
\end{figure}

\textbf{Implication: compute skew translates into stale same-modality exchanges.}
Many decentralized designs implicitly rely on comparable progress rates within an overlay so exchanged updates remain reasonably fresh.
Under modality-induced compute lag, multimodal agents produce shareable same-modality updates less frequently. When peers aggregate the latest available states, these longer refresh intervals can increase update age and slow effective convergence within the modality group~\cite{NEURIPS2022_6db3ea52,pmlr-v238-even24a}. Figure~\ref{fig:staleness_pdf} illustrates this age skew in the GPU setting: from the perspective of a video-only agent, updates from multimodal peers are older than those from video-only peers. Crucially, simply dropping slow contributors can sacrifice useful cross-modal information: these contributors are often the multimodal agents that connect modality-specific learning through local alignment. Under-utilizing their updates can therefore hurt both unimodal and multimodal learning. The challenge is to improve update freshness while preserving these cross-modal contributions.

\textbf{Why we cannot simply remove cross-modal coupling.}
Cross-modal interaction contributes to predictive performance in our evaluated setting. With the same model architecture, data split, and synchronous aggregation schedule, bidirectional interaction achieves 64.5\% multimodal accuracy, compared with 61.8\% for unidirectional interaction and 59.3\% without interaction (Appendix~\ref{app:interaction_ablation}). This motivates preserving the benefit of bidirectional interaction while reducing synchronization-induced waiting.

\textbf{Challenge 1: independent progress with reciprocal interaction.}
Therefore, the computation-level challenge is to reduce synchronization-induced delays and the resulting update staleness
under modality heterogeneity, \emph{while preserving bidirectional cross-modal interaction.}

\subsection{Load-Imbalanced Communication}\label{subsec:motivation_comm}

Decentralized learning performance depends on \emph{who exchanges with whom, how often, and under what bandwidth budget}~\cite{nedic2018network,huang2024overlay}. Under \emph{modality heterogeneity}, this pattern becomes intrinsically imbalanced: multimodal agents exchange multiple modality-specific components and participate in multiple overlays, while unimodal agents participate in only one.
 With comparable exchange rates per overlay and a shared per-agent bandwidth budget, multimodal agents can therefore face higher aggregate communication demand.

\begin{figure}
    \centering
    \includegraphics[width=.7\linewidth]{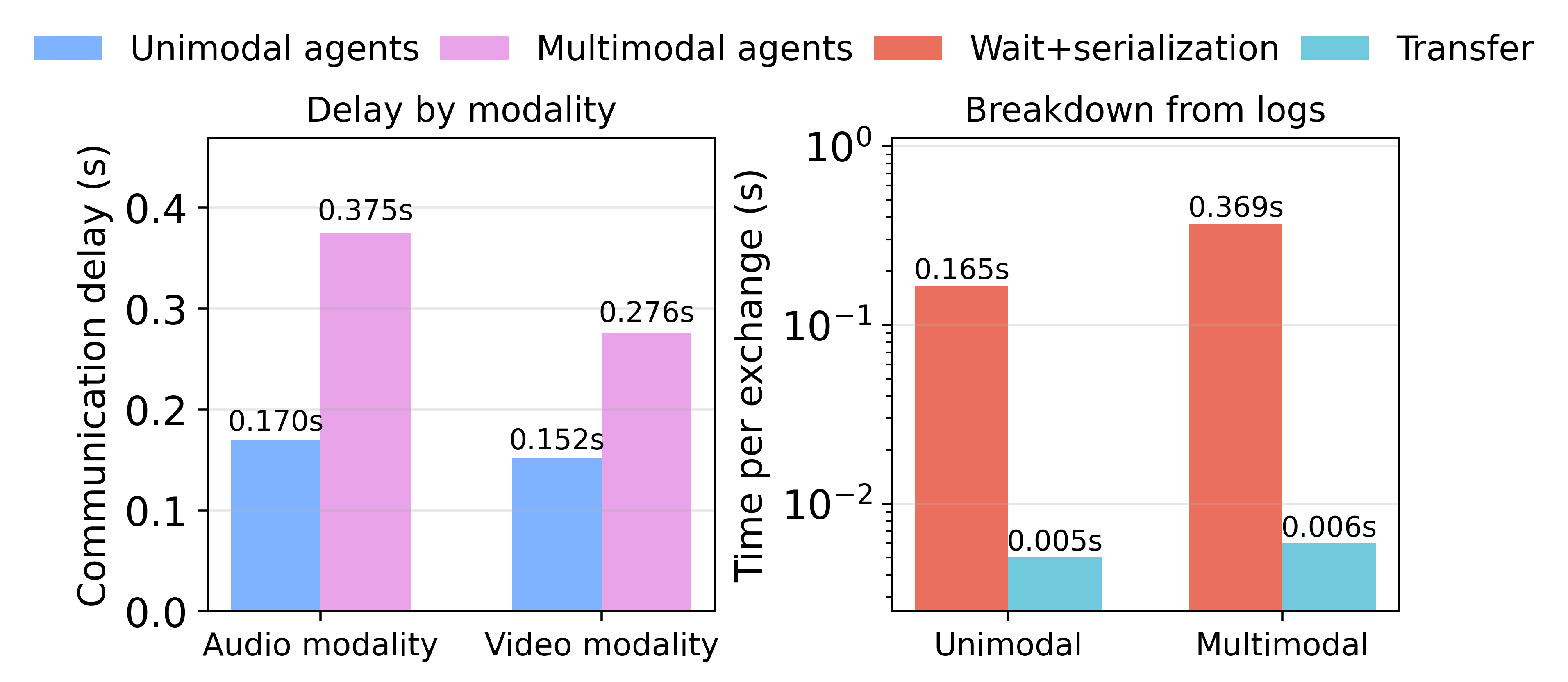}
    \vspace{-0.1in}
    \caption{Transmission delay under a shared 100Mbps full-duplex budget per agent. (Left) per-exchange receive delay within one aggregation window. (Right) completion-time breakdown into enqueue/serialization and bandwidth-limited transfer. }%
    \label{fig:modality_comm}
    \vspace{-0.15in}
\end{figure}

\textbf{Equal bandwidth does not imply equal delay.}
Figure~\ref{fig:modality_comm} isolates this effect with equal 100Mbps full-duplex budgets and the same ring topology for each modality overlay.
Multimodal agents participate in both overlays, so their audio and video exchanges share the same per-agent communication resources. Their reported receive delay is higher in both modalities: 0.375s versus 0.170s for audio ($2.2\times$), and 0.276s versus 0.152s for video ($1.8\times$). Thus, equal per-agent bandwidth budgets can yield substantially different exchange delays when modality memberships differ.

\textbf{Overlapping modality exchanges create contention.}
The delay gap reflects differences in communication load: a multimodal agent concurrently serves and fetches updates across multiple modality overlays, whose transfers share its fixed uplink and downlink budgets. Even with identical per-agent bandwidth limits, these overlapping exchanges compete for the same communication resources. As Figure~\ref{fig:modality_comm}(right) shows, most recorded per-exchange completion time falls in the combined waiting/serialization interval, while the measured transfer interval is smaller. These observations motivate coordinating concurrent exchanges across overlays to control the load placed on multimodal agents.

\textbf{Why both sync and wait-free protocols suffer.}
In synchronous decentralized training, heavily loaded multimodal agents can become bottlenecks and delay aggregation rounds. Wait-free asynchronous training allows local training to continue without waiting for these transfers~\cite{bornstein2023swift}, but the underlying load imbalance remains. Delayed updates can miss local aggregation windows and be incorporated less promptly or less frequently. Thus, asynchrony changes the failure mode rather than removing it: communication delays that previously stalled training can instead leave multimodal contributors underrepresented in aggregation. Because these agents support cross-modal knowledge transfer, reduced participation can weaken information propagation and widen accuracy disparities across agents.

\textbf{Challenge 2: timely exchange under shared bandwidth.}
We therefore need load-aware exchange mechanisms that manage communication load across modality overlays to \emph{reduce hot spots and preserve useful contributions from multimodal agents}, while allowing local training to continue and keeping coordination overhead low under tight bandwidth budgets.

\subsection{Mixing-Staleness-Contention Tension}
\label{subsec:motivation_sched}

The preceding computation and communication imbalances create a coupled peer-selection problem: agents must balance update freshness with broad information mixing under shared bandwidth constraints.
Favoring recent updates can reduce staleness, but repeatedly selecting the same fast peers can underrepresent slower multimodal contributors. Increasing exchange frequency or fan-out to improve mixing, in turn, increases demand on shared links.

\textbf{Why heterogeneity makes the tradeoff dynamic.}
Under modality heterogeneity, the ``best'' peers are neither static nor cheaply observable.
Compute skew (Section~\ref{subsec:motivation_compute}) changes who can produce fresh updates, and multi-overlay load (Section~\ref{subsec:motivation_comm}) changes who can deliver them on time. A peer that is fresh at one moment can become a hot spot the next due to concurrent requests, while multimodal agents may remain systematically slower.
Consequently, fixed schedules that ignore these changes can repeatedly request stale updates or overload popular peers, delaying information diffusion and skewing participation.

\textbf{Why standard aggregation patterns fall short.}
Random peer selection spreads requests across peers in expectation, but does not account for which updates are recent or likely to arrive within the aggregation window. Fixed peer sets or exchange frequencies likewise do not, by themselves, adapt to changes in update availability and serving load. Simple load balancing reduces contention but does not determine which available updates should be mixed. Delayed aggregation reuses late arrivals but does not decide which peers should be contacted next. These mechanisms leave open how to allocate scarce exchanges while preserving both update freshness and the participation of slower multimodal contributors.

\textbf{The practical barrier: contention and probing overhead.}
Even when agents can contact any peer, choosing ``good'' peers is difficult in a wait-free system. If many agents simultaneously target the same high-quality peer, they compete for its limited uplink/downlink, increasing queuing delay and causing transfers to miss aggregation deadlines.
Probing peers for current queue occupancy or training progress adds control messages and can delay selection on high-latency links. This motivates using lightweight shared progress and load metadata to guide peer selection without repeated per-peer probing.

\textbf{Challenge 3: fresh and broad mixing under contention.}
We therefore need a lightweight scheduling mechanism that \emph{balances update freshness with broad participation within a limited exchange budget}, while accounting for serving load and keeping coordination overhead low without active peer probing.

\textbf{\underline{Summary of requirements.}}
These challenges imply three system requirements for multimodal decentralized learning:
\begin{itemize}[leftmargin=1.25em,itemsep=2pt,topsep=2pt]
\item (i) \emph{modality-decoupled progress} without global blocking, while preserving \emph{reciprocal} cross-modal interaction to avoid one-sided alignment (Challenge~1);
\item (ii) \emph{load-aware update exchange} that manages concurrent communication demand across modality overlays and retains useful late arrivals for subsequent aggregation (Challenge~2);
\item (iii) \emph{contention-aware peer selection} that balances freshness and broad participation within the available exchange budget, with low control overhead (Challenge~3).
\end{itemize}
These requirements are coupled: local modality progress determines which updates are available, communication contention affects which arrive in time, and peer selection determines whose information enters aggregation. \MODAIC coordinates these decisions within a single runtime.

\section{\MODAIC Design}

\begin{figure}[t]
    \centering
    \includegraphics[width=.75\linewidth]{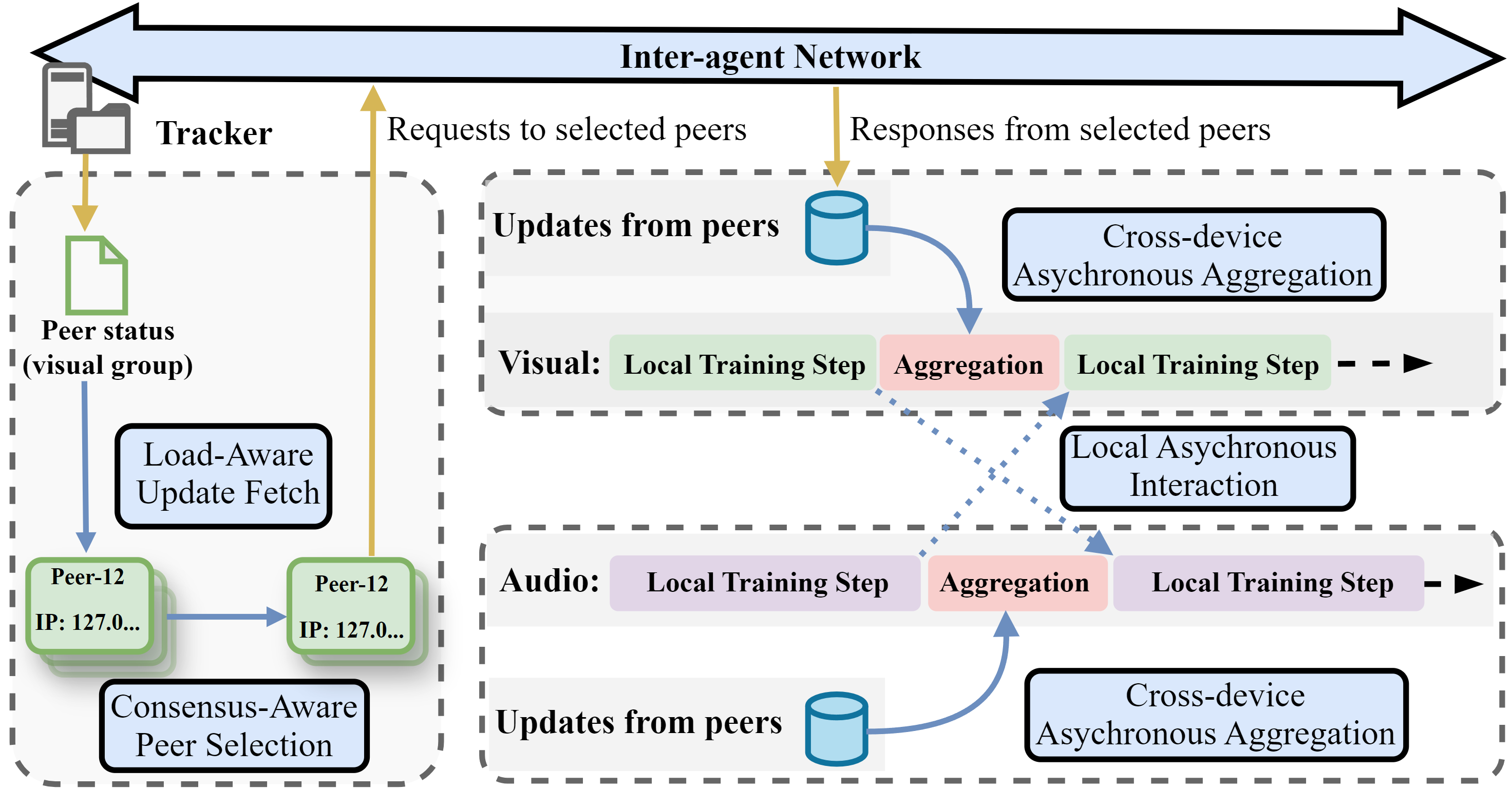}
    \vspace{-0.1in}
       \caption{\MODAIC overview on a multimodal (audio+visual) agent. The tracker is control-plane only: it maintains admission/log metadata and is not on the model-data path; updates are exchanged P2P. Each modality runs a wait-free train/pull/serve pipeline with cached peer updates and asynchronous aggregation, while local interaction enforces reciprocal cross-modal feature exchange.}
    \label{fig:sys_overview}
      \vspace{-0.15in}
\end{figure}

To meet the requirements distilled in Section~\ref{sec:motivation}, we present \MODAIC
(\underline{MO}dality-\underline{D}ecoupled \underline{A}synchrony with \underline{I}nterleaved \underline{C}oordination), a wait-free runtime for multimodal decentralized learning under heterogeneous compute and bandwidth. \MODAIC follows one guiding principle: \emph{decouple modality progress while coordinating reciprocal local interaction and bandwidth-limited peer-update exchange.}
Here, wait-free means that local training proceeds without waiting for partner features or remote model transfers. Figure~\ref{fig:sys_overview} summarizes the per-agent runtime.

This principle leads to three design requirements. First, local cross-modal interaction must remain \emph{reciprocal}:
each modality must contribute matched features before consuming its partner's features.
Second, update exchange must be \emph{deadline-resilient}:
aggregation proceeds with available updates, while useful late arrivals remain eligible for subsequent windows.
Third, peer selection must account for \emph{update availability, participation, and serving load}, so scarce exchanges support both timely updates and broad participation without active peer probing.

\MODAIC meets these requirements through three integrated components. \textbf{\textit{Hierarchical Asynchronous Collaboration}} decouples training and aggregation across modalities while preserving reciprocal feature exchange (Section~\ref{subsec:hier_async}). \textbf{\textit{Load-Aware Update Fetching}} overlaps computation and communication, limits concurrent serving load through lightweight admission control, and retains late arrivals in per-modality caches (Section~\ref{subsec:selective_peer_design}). \textbf{\textit{Consensus-Aware Peer Selection}} uses metadata about update availability, recent participation, and loss similarity to guide peer exchange (Section~\ref{subsec:peer_selection_design}). These components coordinate coupled decisions: local asynchrony determines when updates become available; admission control and caching determine which exchanges can proceed and which completed updates remain available; and peer selection determines which contributions to request and incorporate. A centralized metadata tracker supports coordination, while all model states are exchanged directly between agents without centralized aggregation or model-data relay.

\subsection{Hierarchical Asynchronous Collaboration}\label{subsec:hier_async}

This component addresses Challenge~1 (Section~\ref{subsec:motivation_compute}): reduce compute-induced staleness by allowing each modality to progress without global blocking, while still preserving the \emph{bidirectional} cross-modal interaction required for multimodal accuracy. \MODAIC coordinates asynchrony at two levels: (i) \emph{local} asynchronous interaction between modalities within an agent, and (ii) \emph{cross-device} asynchronous aggregation within each modality overlay.

\begin{figure*}[t]
    \centering
    \includegraphics[width=.75\linewidth]{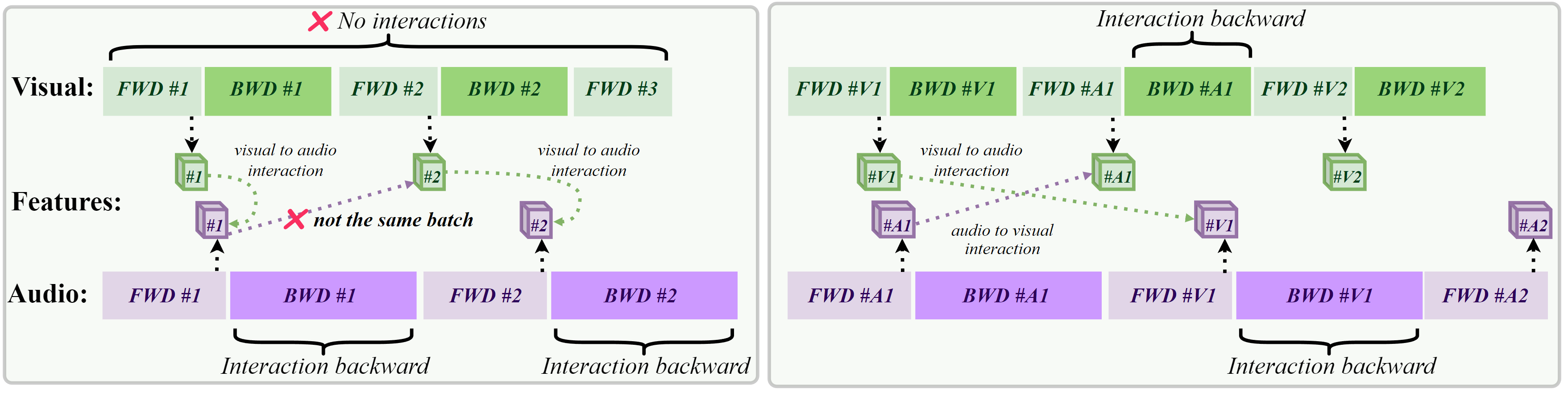}
    \vspace{-0.1in}
       \caption{How MODAIC preserves reciprocal interaction. (Left) With a shared data stream, the faster modality repeatedly reaches interaction points first, consuming cached partner features more often than it contributes; cached features may also mismatch the partner’s current batch. (Right) MODAIC uses per-modality paired streams and a contribute-before-consume rule to ensure matched inputs and bidirectional exchange under asynchronous execution.}
    \label{fig:sync_train}
      \vspace{-0.15in}
\end{figure*}

\ding{182} \textbf{Local asynchronous interaction: preserving reciprocity under asynchrony.}
A naive design is to run modalities independently
 and exchange features whenever they become available. Under modality heterogeneity and without coordination, this quickly becomes imbalanced: the faster modality repeatedly reaches interaction points first and mostly \emph{consumes} the partner’s cached features, while the slower modality mostly \emph{produces} them. Moreover, if both modalities draw from a single shared data stream, cached features may correspond to a different data sample than the partner input used later for interaction, biasing alignment, as illustrated in Figure~\ref{fig:sync_train} (left).

\MODAIC prevents this collapse with two simple rules as illustrated in Figure~\ref{fig:sync_train} (right).
\emph{(R1) Independent streams with matched interaction inputs:} each modality advances its own stream of paired multimodal batches $(x_m,x_o,y)$. For interaction, it uses the matched counterpart input stored with the partner's cached features. \emph{(R2) Contribute-before-consume:} a modality may consume its partner's features only after contributing its own features for the partner. R1 preserves sample correspondence, while R2 preserves reciprocal exchange.

We implement (R2) using a lightweight \textit{credit-and-buffer} mechanism. Each modality $m$ maintains a single-slot buffer $\textsc{Buf}[m]$ that stores $(x_o,y,h_m)$ from the most recent paired sample it processed, and a boolean credit flag $\textsc{Credit}[m]$. When modality $m$ produces batch feature $h_m$, it fills $\textsc{Buf}[m]$ (if empty) and sets $\textsc{Credit}[m]$ to $\texttt{true}$. An interaction step can occur only when $\textsc{Credit}[m]$ is $\texttt{true}$ \emph{and} the partner buffer is non-empty.
Then modality $m$ consumes the partner tuple $(x_m,y,h_o)$ from $\textsc{Buf}[o]$, computes $h_m$ on the stored input $x_m$ from the same paired sample, and applies the interaction loss.
The single-slot design bounds memory and avoids accumulating stale feature queues. The full procedure is given in Appendix~\ref{app:async_interaction}.

\ding{183} \textbf{Cross-device asynchronous aggregation: per-modality progress without global barriers.}
Beyond local interaction, \MODAIC supports wait-free, modality-level sharing across agents.
Each modality $m$ runs an independent aggregation loop: it issues asynchronous pulls for same-modality updates, continues local training, and at each aggregation deadline mixes the arrived or cached updates for modality $m$.
Neither cross-device aggregation nor local feature exchange requires modalities to reach a common training step. This allows each modality to refresh its parameters as its local computation and available peer updates permit, without waiting for slower modalities or remote transfers.

The runtime abstraction is not hard-coded to two modalities. For an agent with modality set $\mathcal{M}_i$, \MODAIC instantiates one train/pull/serve pipeline per modality and one sharing overlay per modality.  Local interaction can be pairwise or fusion-based, with the \emph{contribute-before-consume} rule requiring that consumed features come only from modalities that have contributed in the current interaction window. Thus, the control/data-plane separation and load-aware fetching design remain unchanged, although local interaction state and overlay membership grow with $|\mathcal{M}_i|$.

\subsection{Load-Aware Update Fetching} \label{subsec:selective_peer_design}

This component addresses Challenge~2 (Section~\ref{subsec:motivation_comm}) by making peer-to-peer update exchange efficient under bandwidth limits, so multimodal agents are neither persistent hot spots nor systematically under-aggregated. \MODAIC provides a load-aware runtime that (i) overlaps training with communication, (ii) limits inbound load via lightweight admission control, and (iii) reuses late-but-useful arrivals via caching instead of discarding them.
This layer defines the \emph{feasible} set of transfers (which peers are admissible and which updates are locally available) that our consensus-aware peer selection builds on (Section~\ref{subsec:peer_selection_design}).

\begin{figure}[t]
    \centering
    \includegraphics[width=.75\linewidth]{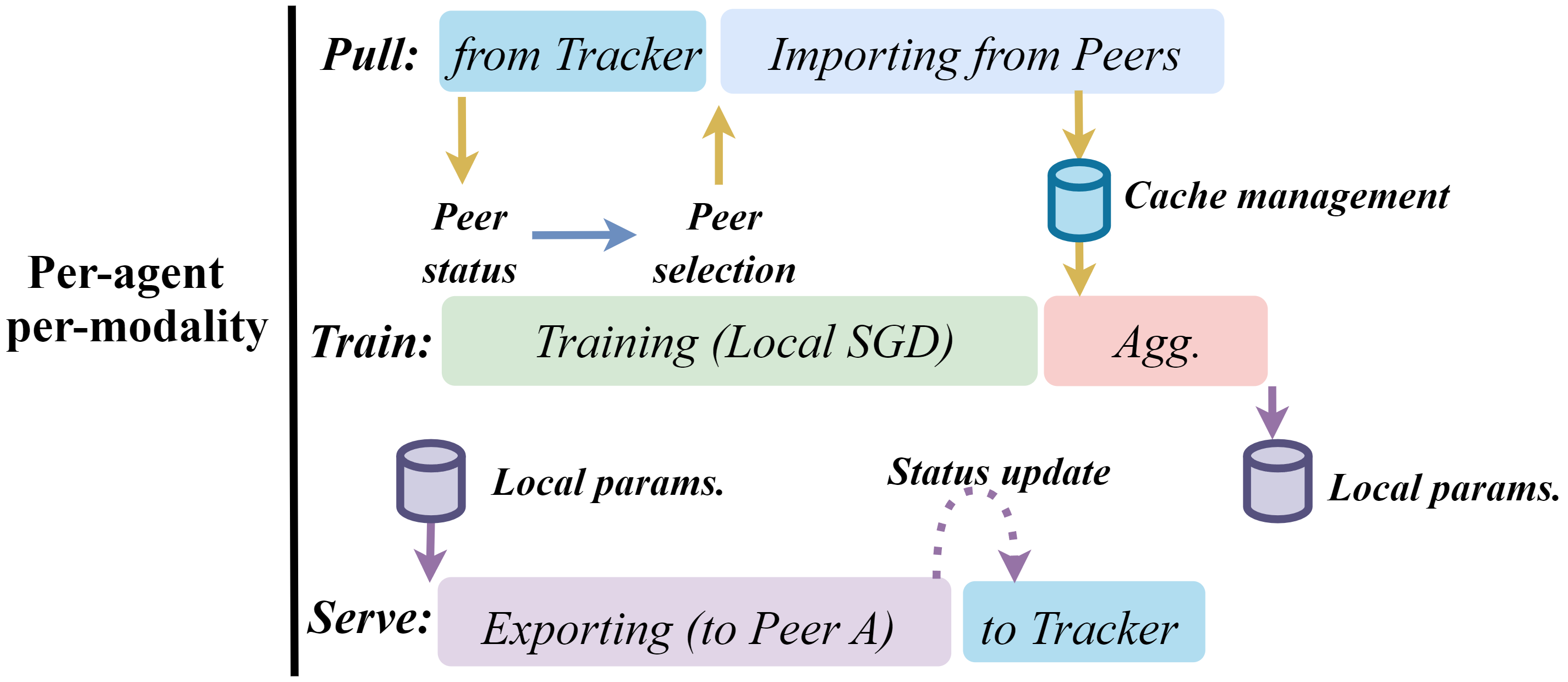}
   \vspace{-0.1in}
      \caption{Load-aware update fetching runtime. Each modality runs concurrent \textit{Train}, \textit{Pull}, and \textit{Serve} workers. Pulls reserve admissible peers through the metadata tracker, model states are exchanged P2P, and aggregation consumes arrived or cached updates without blocking training.}
    \vspace{-0.15in}
    \label{fig:concurrent}
\end{figure}

\ding{182} \textbf{Communication-computation concurrency.}
As shown in Figure~\ref{fig:concurrent}, each agent runs a per-modality, wait-free pipeline with three concurrent routines. \textit{Train} continuously advances local parameters via SGD. \textit{Serve} replies to inbound pull requests with the latest locally available same-modality state. \textit{Pull} asynchronously requests peer updates for the next aggregation event.  Aggregation executes at a configurable cadence, such as after a local training window, and mixes whichever updates have arrived by the deadline; training is never blocked by communication. Because these routines are modality-isolated, a fast modality can continue training and exchanging updates without waiting for slower modalities on the same agent.

\ding{183} \textbf{Load-balanced admission control.}
Unrestricted pulling can create many-to-one contention when agents target the same peer. To limit concurrent serving load without active peer probing, the tracker maintains a count $a_j^m$ of reserved or active serving slots and a configured capacity $c_j^m$ for each peer $j$ and modality $m$. Peers with $a_j^m<c_j^m$ are eligible to serve additional requests. Before a pull starts, one slot must be atomically reserved by incrementing $a_j^m$, and the slot is released when the transfer completes or is terminated. An aggregation deadline alone does not terminate a pending transfer, since its result may still be useful in a later window.

This admission mechanism limits the concentration of concurrent requests at individual peers. It complements peer selection: selection identifies desirable contributions, while admission controls whether the corresponding transfers can proceed. Per-modality capacities bound serving concurrency, while their configuration must account for the bandwidth shared by an agent's modality overlays.

\ding{184} \textbf{Caching late arrivals.}
Aggregation does not wait for in-flight transfers: at each deadline, an agent mixes the updates that have arrived. Unlike designs that cancel or discard late transfers,
\MODAIC stores completed arrivals in a small per-modality receive cache $\mathcal{B}_i^m$. Late updates that miss the current window remain eligible for the next aggregation event, turning deadlines into a \emph{soft boundary} that determines \emph{when} an update is applied rather than whether transmitted bytes are discarded. Cached entries are versioned and later filtered/reweighted by the selection policy, so caching salvages bandwidth without forcing the agent to apply every late update. We bound cache size with an age-based eviction policy; details are in Appendix~\ref{app:cache_details}. Caching preserves opportunities to reuse completed transfers without requiring every late arrival to be aggregated.

\subsection{Consensus-Aware Peer Selection}\label{subsec:peer_selection_design}

Built on the load-aware update fetching layer (Section~\ref{subsec:selective_peer_design}), this component addresses Challenge~3 (Section~\ref{subsec:motivation_sched}): improve mixing while controlling staleness and limiting many-to-one contention \emph{without} active peer probing. Given the admissible peers and locally available (possibly cached) updates, \MODAIC uses a lightweight selection policy that balances \textit{freshness}, \textit{participation fairness}, and \textit{loss similarity}. The fetching layer controls concurrent serving load, while the selection policy determines which admissible peers to contact and which available updates to incorporate.

\ding{182} \textbf{Coordination log and selection principles.}
To achieve consensus-aware mixing without costly active probing, \MODAIC uses a lightweight coordination-plane signal: an \emph{append-only coordination log} maintained by a tracker that is \emph{not} on the model-data path.
After agent $j$ completes an aggregation window for modality $m$ at time $t$, it appends
\begin{align*}
(\text{cid-}j, m, t, \ell_j^m, a_j^m, c_j^m, \text{IP}) \in \mathcal{J},
\end{align*}
where $\ell_j^m$ is a cheap pace signal (e.g., loss), $(a_j^m,c_j^m)$ are the current active-transfer count and capacity already used by admission control (Section~\ref{subsec:selective_peer_design}), and $\text{IP}$ is the IP-address of the agent.

The log plays two roles: it exposes \emph{update availability} (who has a recent admissible update) and supports peer selection guided by three principles: \textit{freshness} (prefer unaggregated updates), \textit{fairness} (avoid repeatedly sampling the same peers), and \textit{pace similarity} (prefer peers at similar training state to reduce drift). The tracker stores only metadata; all model states are fetched directly between agents. Agents therefore make decisions by reading a small log snapshot and then initiating P2P transfers, without direct peer state queries. If the tracker snapshot is stale or temporarily unavailable, agents fall back to the last admissible snapshot and cached updates; this may reduce peer freshness but does not block local training.

\ding{183} \textbf{Selection scores.}
Peer selection is performed independently per modality. At peer-selection time $t$ for modality $m$,
agent $i$ reads the coordination-log snapshot for the current modality overlay and constructs an exclude set $\mathcal{E}^m_i(t)$ containing (i) itself and (ii) the (peer,version) update IDs already aggregated in the previous window. It then builds a fresh candidate pool by keeping the latest eligible entry per peer:
\begin{align}
\mathcal{C}^m_i(t)\!=\!\Big\{\text{latest entry per peer } j\neq i \text{ from }\mathcal{J}\Big|j\notin\mathcal{E}^m_i(t)\Big\}.
\end{align}
For each candidate $j\in\mathcal{C}^m_i(t)$, we compute a fairness-similarity score
\begin{align}
s^m_{ij} &=(1-\alpha)\cdot\underbrace{\frac{1}{1+\mathrm{count}^m_{ij}}}_{\text{fairness}} +\alpha\cdot\underbrace{\exp\!\Big(-|\ell^m_i-\ell^m_j|/\beta\Big)}_{\text{pace similarity}},
\label{eq:score}\allowdisplaybreaks\\[-2pt]
p^m_{ij} &= \frac{s^m_{ij}}{\sum_{u\in\mathcal{C}^m_i(t)} s^m_{iu}},
\label{eq:prob}
\end{align}
where $\mathrm{count}_{ij}$ is the number of times $i$ has recently aggregated from $j$, $\alpha\in[0,1]$ trades off fairness vs. similarity, and $\beta>0$ is a temperature. Agent $i$ then samples up to $K$ peers \emph{without replacement} according to $\{p^m_{ij}\}$. Let $S^m_i(t)$ denote the sampled set (and write $S^m_i$ when $t$ is clear).

\ding{184} \textbf{Aggregation weights: optional loss-similarity reweighting.}
At the aggregation deadline ($t+\Delta$), given the per-modality receive cache $\mathcal{B}_i^m$ (of modality $m$), agent $i$ aggregates the received updates using base weights $W^m_{ij}$ (uniform by default). Optionally, we apply a lightweight loss-similarity reweighting to prioritize pace-compatible peers:
\begin{equation}
\tilde{w}^m_{ij}\propto W^m_{ij}\exp\!\big(-|\ell^m_i-\ell^m_j|/\beta_w\big),
\label{eq:reweight}
\end{equation}
where $\beta_w>0$ controls how sharply we favor similar-loss peers (larger $\beta_w$ is closer to uniform). To avoid overmixing, we renormalize $\tilde{w}^m_{ij}$ over all updates.

\ding{185} \textbf{Fallback for fast agents.}
Fast agents may reach an aggregation deadline before enough \emph{fresh, unaggregated} peer updates are available (e.g., under heterogeneous compute or bandwidth), which can cause drift if they repeatedly mix from an undersized pool. To maintain stability without blocking, we use \emph{fallback selection}: if the fresh candidate pool is small, we fill the remaining peer budget using locally cached (already-downloaded) updates that are still pace-compatible.

Concretely, if $|\mathcal{B}^m_i| = N < K$, agent $i$ first samples $N$ updates from $\mathcal{B}^m_i$ (with same selection scores), and then samples the remaining $K-N$ peers from the aggregated-cache pool $\mathcal{A}^m_i$ ($\mathcal{A}^m_i$ stores previously aggregated updates). Fallback sampling uses the same fairness-similarity score in Eq.~\ref{eq:score} (restricted to $\mathcal{A}^m_i(t)$), so it preserves exploration (fairness) while prioritizing pace similarity.
Because $\mathcal{A}^m_i(t)$ is on-device, this step incurs no additional network overhead.

\textbf{\underline{Overall.}}
At each modality-specific aggregation cycle, agent $i$ reads the latest log snapshot, obtains admissible peers through the load aware fetching layer, samples up to $K$ peers using the scores in Eq.~\ref{eq:score}, and issues asynchronous pulls subject to admission control. At the local aggregation deadline, it selects eligible completed updates from $\mathcal{B}_i^m$ and supplements them from $\mathcal{A}_i^m$ when needed and available, up to the aggregation budget. It then combines the selected updates with its local modality model using normalized weights, with optional loss-similarity reweighting (Eq.~\ref{eq:reweight}). Pending transfers do not block aggregation, and eligible late arrivals remain available for subsequent cycles. The pseudocode is given in Appendix~\ref{app:peer_selection}.

\section{System Implementation}\label{sec:implementation}

We implement \MODAIC as a containerized multi-agent training framework in \texttt{Python3} (about 3.2K lines of code) on top of \texttt{PyTorch}. \MODAIC exposes a small set of APIs for (de)serializing modality-specific model states and for scheduling per-modality training and communication pipelines. This design enables cross-agent modality sharing over commodity IP networks using TCP connections, without requiring dedicated links or a centralized model-data relay.

\textbf{Containerized agents.}
Each agent runs in an isolated Docker container built from a \texttt{PyTorch CUDA} runtime image. Containers encapsulate the full training stack, including model components, optimizer, dataloaders, caches, and communication workers, and provide fault isolation. Each agent maintains modality-specific components (encoders and task heads) and executes \MODAIC via per-modality pipelines that overlap local training with communication, consistent with the design in Section~\ref{subsec:selective_peer_design}. For each exchange, we serialize only the required modality-specific state together with small metadata (peer id, modality, version, timestamp, and loss/progress signal). This keeps P2P model-transfer payloads scoped to the relevant modality and avoids device-wide synchronization across unrelated modalities.

\textbf{GPU sharing and isolation.}
To support many concurrent agents on a single GPU server while limiting interference, we enable NVIDIA Multi-Process Service (MPS) to partition GPU execution across containers. Each container is assigned a controlled GPU share (e.g., via MPS active-thread percentage), preventing any single agent from monopolizing the device. Within each container, \MODAIC implements the per-modality \textit{pull}, \textit{serve}, and \textit{train} routines as concurrent workers, so communication proceeds while local SGD continues and no global barrier is introduced across modalities. To avoid unintended cross-modality blocking on shared accelerators, each modality executes forward/backward kernels on its own CUDA stream, while a lightweight mutex guards only optimizer steps and parameter serialization to prevent concurrent writes to shared tensors. Received peer updates are stored in per-modality caches to tolerate stragglers and bursty links, and to enable communication-free reuse of late arrivals in subsequent aggregation windows.

\textbf{Control plane and peer-to-peer data path.}
Consistent with Sections~\ref{subsec:selective_peer_design} and~\ref{subsec:peer_selection_design}, each agent maintains a control-plane connection to a lightweight tracker service that records only small status metadata for admission control and log-guided peer selection. The tracker is not on the model-data path (i.e., it is not a parameter server): all model updates are transmitted directly between peers over the data plane. All tracker measurements use the same single-process Python tracker used in the training experiments. If the tracker is temporarily unreachable, agents continue wait-free training and aggregation using their most recent admissible-peer snapshot and locally cached updates until connectivity resumes. The tracker can be replicated or sharded by modality for availability; we quantify its overhead and failure behavior in Section~\ref{sec:tracker_eval}.

\textbf{Network emulation and deployment.}
For controlled local experiments, we construct virtual overlay networks among containers using Mininet~\cite{mininet}. Mininet instantiates virtual links and allows us to configure per-agent bandwidth and latency. We attach all agent containers to a virtual switching fabric, assign each container a static IP address for stable addressing, and enforce heterogeneous link budgets matching our experimental settings. Unless stated otherwise, local experiments do not inject random packet loss; WAN-style RTT variation and loss stress tests, as well as the cross-region cloud deployment, are evaluated separately in Section~\ref{sec:exp}.

\section{Evaluation}\label{sec:exp}

We evaluate \MODAIC on a controlled local testbed and a cross-region cloud deployment.
Building on the resource effects characterized in Section~\ref{sec:motivation}, we examine whether coordinating local cross-modal interaction and peer-update exchange improves learning under heterogeneous compute and limited bandwidth. We emphasize both average and worst-agent accuracy, then examine the contributions of the runtime components and the cost of coordination.
Concretely, our evaluation answers:
\begin{itemize}[leftmargin=1.25em,itemsep=2pt,topsep=2pt]
\item \textbf{End-to-end benefit:} Does \MODAIC improve both average and worst-agent accuracy for unimodal and multimodal agents, and does it reduce the bandwidth needed to reach a target worst-agent accuracy (Section~\ref{sec:main_results})?
\item \textbf{Are simpler fixes enough?}
How much of the benefit can load-balanced random selection and delayed-update reuse recover (Section~\ref{sec:simple_variants})?
\item \textbf{Where do the gains come from?} How do reciprocal local interaction, load-aware update fetching, and consensus-aware peer selection contribute, individually and in the tested combinations
(Section~\ref{sec:component_synergy})?
\item \textbf{Control-plane practicality:}
What overhead does the metadata tracker incur, and how does training behave under delayed or partially unavailable tracker access (Section~\ref{sec:tracker_eval})?
\item \textbf{Robustness under heterogeneity:}
How do bandwidth, stragglers, system scale, and WAN-like network conditions affect performance (Section~\ref{sec:sensitivity})?
\item \textbf{Wide-area deployment:} Do the same trends hold in a real cross-region cloud testbed, and does \MODAIC reduce time-to-accuracy under CPU-driven training and heterogeneous wide-area links (Section~\ref{sec:cloud_exp})?
\end{itemize}

\subsection{Experimental Setup}\label{sec:setup}

Our evaluation includes both a controlled local testbed and cloud-based experiments. Both use the same containerized \MODAIC implementation and differ only in their network configuration. The local testbed provides GPU-based training and controlled network conditions, whereas the cloud deployment uses CPU-only training over cross-region links. We present the main results and most controlled analyses on the local testbed, and in Section~\ref{sec:cloud_exp} we reproduce a subset of the results in the cloud to demonstrate practicality under real wide-area links. Importantly, the local experiments are not simulations of training: all methods run as containerized PyTorch agents, while Mininet controls communication among containers.

\textbf{Testbed and default configuration.}
All local testbed experiments run on a server with an AMD Ryzen Threadripper PRO 7955WX CPU and three NVIDIA RTX PRO 5000 GPUs. GPU execution is shared across containers using NVIDIA MPS, as described in Section~\ref{sec:implementation}.
The default configuration has 30 agents: 10 hold both modalities, and 20 hold exactly one modality, split evenly between the two modalities. Unless otherwise stated, the configured bandwidth budget is 200Mbps per agent, and we measure local-testbed accuracy at 1000s.

\textbf{Datasets and Non-IID setting.}
We evaluate on two multimodal benchmarks that cover both lightweight sensor modalities and compute-heavier audio-visual encoders.
(1) \text{KU-HAR}~\cite{kuhar} is a human activity recognition dataset with two sensor modalities, accelerometer and gyroscope. We use an 8-class subset (walking, walking upstairs, walking downstairs, sitting, standing, laying, jumping, and running), and adopt ResNet-9~\cite{he2016resnet} as the backbone for both modalities.
(2) \text{AVE}~\cite{avedataset} (Audio-Visual Event) contains 10-second video clips from 28 event categories, with synchronized audio and visual streams. We use ResNet-18~\cite{he2016resnet} to encode audio and MobileNetV3-Large~\cite{MobileNet} to encode visual inputs.
To model non-IID data, we assign each agent a label distribution sampled from a Dirichlet distribution with concentration parameter $\alpha$~\cite{hsu2019measuring}. Unless otherwise noted, we report results with $\alpha=0.5$, and we vary $\alpha$ in additional studies (Appendix~\ref{app:extra_results}).

\textbf{Training configuration.}
We use SGD with learning rate $10^{-3}$, momentum 0.9, weight decay $5\times10^{-4}$, and batch size 32. The local training window contains one SGD step. Model and method-specific configuration details are given in Appendix~\ref{app:impl_details}.

\textbf{Baselines.} We compare against four asynchronous decentralized learning baselines: AD-PSGD~\cite{ad-psgd}, AsyDFL~\cite{asydfl}, Swift~\cite{bornstein2023swift}, and DivShare~\cite{biswas2025boosting}. AD-PSGD assigns nodes to active/passive roles, and active nodes push updates to passive ones to avoid synchronization deadlocks. Swift performs peer-to-peer exchanges with a consensus-based neighbor rule. AsyDFL combines asynchronous update pushing with resource-aware peer selection. DivShare improves communication efficiency by slicing model parameters and sharing only subsets in each exchange. In addition, we include representative synchronous baselines in our main results: Harmony~\cite{ouyang2023harmony}, DMML-KD~\cite{kd-dfl}, and FedLay~\cite{hua2024towards}. Harmony is a two-stage multimodal federated learning scheme, DMML-KD enables cross-modality transfer via feature generation and distillation, and FedLay couples a practical overlay topology with a model-exchange protocol to mitigate data and client heterogeneity. We adapt Harmony to our peer-to-peer setting by replacing server aggregation with overlay-based peer exchange while keeping its core interaction mechanisms unchanged; detailed adaptations and hyperparameter matching are provided in Appendix~\ref{app:impl_details}.

\textbf{Metrics.}
We report both average and worst-agent accuracy. For agents with a single modality, U-Avg and U-Worst denote the average and worst accuracy among unimodal agents. For agents with multiple modalities, M-Avg and M-Worst denote the average and worst accuracy among multimodal agents. Worst-agent metrics are central because modality heterogeneity first harms agents that are slow, overloaded, or under-aggregated. We also report the average-worst gap to characterize within-group accuracy disparity, and use wall-clock learning curves and time-to-accuracy to assess training efficiency. Accuracy differences are reported in percentage points.

\begin{table*}[t]
\centering
\providecommand{\seedstd}[1]{{\scriptsize$\pm$#1}}
\setlength{\tabcolsep}{6.5pt}
\renewcommand{\arraystretch}{1.15}
\scalebox{0.8}{
\begin{tabular}{lcccc|cccc}
\toprule
\multirow{2}{*}{\textbf{Method}} &
\multicolumn{4}{c}{\textbf{AVE}} &
\multicolumn{4}{c}{\textbf{KU-HAR}} \\
\cmidrule(lr){2-5}\cmidrule(lr){6-9}
& \textbf{U-Avg} & \textbf{U-Worst} & \textbf{M-Avg} & \textbf{M-Worst}
& \textbf{U-Avg} & \textbf{U-Worst} & \textbf{M-Avg} & \textbf{M-Worst} \\
\midrule
Harmony   & 30.50\seedstd{1.08} & 27.44\seedstd{2.17} & 42.96\seedstd{1.35} & 37.75\seedstd{2.52} & 65.11\seedstd{0.52} & 62.01\seedstd{1.17} & 84.38\seedstd{0.43} & 78.27\seedstd{1.56} \\
DMML-KD   & 33.90\seedstd{1.22} & 32.17\seedstd{1.89} & 45.17\seedstd{0.82} & 40.67\seedstd{1.85} & 66.94\seedstd{0.42} & 65.58\seedstd{1.35} & 86.70\seedstd{0.70} & 80.80\seedstd{0.90} \\
FedLay    & 32.77\seedstd{1.04} & 27.68\seedstd{1.55} & 44.91\seedstd{0.76} & 37.01\seedstd{1.97} & 67.89\seedstd{0.70} & 65.04\seedstd{0.90} & 85.20\seedstd{0.48} & 79.41\seedstd{1.72} \\
\midrule
AD-PSGD   & 25.71\seedstd{1.36} & 19.83\seedstd{1.21} & 28.35\seedstd{1.18} & 20.14\seedstd{1.47} & 64.76\seedstd{0.80} & 56.48\seedstd{1.12} & 77.66\seedstd{0.86} & 75.06\seedstd{1.33} \\
AsyDFL    & \heatmaplevel{3}{37.42\seedstd{1.04}} & 26.60\seedstd{5.11} & \heatmaplevel{3}{52.62\seedstd{1.17}} & \heatmaplevel{3}{45.27\seedstd{3.39}} & \heatmaplevel{3}{71.11\seedstd{0.83}} & \heatmaplevel{3}{62.71\seedstd{2.33}} & \heatmaplevel{3}{86.57\seedstd{1.02}} & \heatmaplevel{3}{79.31\seedstd{2.27}} \\
Swift     & 33.30\seedstd{2.94} & 19.18\seedstd{7.51} & 50.16\seedstd{2.48} & 41.24\seedstd{6.88} & 69.81\seedstd{1.49} & 60.33\seedstd{3.83} & 83.60\seedstd{1.76} & 77.40\seedstd{4.28} \\
DivShare  & 33.65\seedstd{2.95} & \heatmaplevel{3}{27.17\seedstd{5.69}} & 51.06\seedstd{2.64} & 38.29\seedstd{5.68} & 69.95\seedstd{0.95} & 61.83\seedstd{3.91} & 85.01\seedstd{1.60} & 78.98\seedstd{4.69} \\
\MODAIC   & \heatmaplevel{6}{\textbf{39.57}\seedstd{1.51}} & \heatmaplevel{6}{\textbf{35.16}\seedstd{2.80}} & \heatmaplevel{6}{\textbf{59.74}\seedstd{1.24}} & \heatmaplevel{6}{\textbf{56.28}\seedstd{2.23}} & \heatmaplevel{6}{\textbf{73.24}\seedstd{1.26}} & \heatmaplevel{6}{\textbf{64.92}\seedstd{1.91}} & \heatmaplevel{6}{\textbf{87.92}\seedstd{1.19}} & \heatmaplevel{6}{\textbf{81.56}\seedstd{2.48}} \\
\bottomrule
\end{tabular}}
\caption{\textbf{End-to-end accuracy under the default setting.} We report unimodal and multimodal accuracy as U-Avg/U-Worst and M-Avg/M-Worst (average and worst across unimodal or multimodal agents). Entries are mean$\pm$std over 5 seeds. \MODAIC improves both average accuracy and worst-agent performance.}
\label{tab:default_200mbps_dir_datasets}
\vspace{-0.25in}
\end{table*}

\subsection{End-to-End Results} \label{sec:main_results}

\textbf{Accuracy and worst-agent fairness.}
Table~\ref{tab:default_200mbps_dir_datasets} shows that \MODAIC achieves the best unimodal and multimodal accuracy on both benchmarks, and more importantly, markedly improves \emph{worst-agent} performance.  On AVE, \MODAIC reaches 39.57 (U-Avg)/35.16 (U-Worst) and 59.74 (M-Avg)/56.28 (M-Worst), improving the strongest asynchronous baseline by +2.15 (U-Avg), +7.99 (U-Worst), +7.12 (M-Avg), and +11.01 (M-Worst). The average-worst gaps shrink to 4.41 (unimodal) and 3.46 (multimodal), compared to $\ge$5.88 and $\ge$7.35 for asynchronous baselines, indicating substantially better cross-agent fairness under modality heterogeneity. Similar trends hold on KU-HAR: \MODAIC raises U-Worst and M-Worst while keeping the highest averages, suggesting gains are broadly shared rather than concentrated on a few fast agents.

These improvements target the failure mode in Section~\ref{sec:motivation}: under modality heterogeneity and bandwidth limits, slow or multimodal bridge agents are first to be under-aggregated or to incorporate stale information as transfers queue and miss aggregation windows. \MODAIC counteracts this with deadline-resilient, contention-aware exchange (pull-based fetching with admission control and late-arrival caching), modality-decoupled asynchrony, and consensus-aware peer selection that avoids repeatedly mixing from over-targeted peers.

\begin{table}[t]
\centering
\setlength{\tabcolsep}{5pt}
\begin{tabular}{lr}
\toprule
\textbf{Method} & \textbf{Required Bandwidth} \\
\midrule
AD-PSGD        & 1000--1200 Mbps \\
AsyDFL         & 450--500 Mbps\\
Swift          & 400--500 Mbps\\
DivShare          & 300--400 Mbps\\
\MODAIC & \textbf{150--250} Mbps\\
\bottomrule
\end{tabular}
\caption{\textbf{Bandwidth-to-target on AVE.} Per-agent bandwidth  required for each method to reach \textbf{M-Avg}=60 at 1000s. Lower is better.}
\label{tab:bandwidth_for_60_multiworst}
\vspace{-0.25in}
\end{table}

\textbf{Bandwidth-to-target efficiency.}
Beyond improving accuracy at a fixed budget, \MODAIC reaches the same worst-agent target with substantially less bandwidth. Table~\ref{tab:bandwidth_for_60_multiworst} reports the per-agent bandwidth required to reach M-Avg$=60$ at 1000s on AVE: AD-PSGD needs 1000-1200Mbps, AsyDFL 450-500Mbps, Swift 400-500Mbps, and DivShare 300-400Mbps, whereas \MODAIC achieves the same target with only 150-250Mbps. This up to 2.7$\times$ reduction in bandwidth-to-target makes multimodal decentralized training considerably more deployable over bandwidth-limited inter-site links, aligning with the WAN-style constraints that motivate our design. We next test whether simpler fixes can recover these gains.

\textbf{Final accuracy ceiling.}
We also check whether the wait-free design improves time-to-accuracy at the cost of eventual accuracy. In a longer run until saturation, \MODAIC reaches 69.8\% final M-Avg, compared with 68.1\% for \textsc{DMML-KD}, the strongest synchronous baseline in this setting (Appendix~\ref{sec:convergence_app}). Thus, \MODAIC's gains are not merely early-time artifacts; the runtime improves bandwidth efficiency and time-to-accuracy without lowering the final accuracy ceiling.

\subsection{Are Simpler Fixes Enough?}\label{sec:simple_variants}

A key question is whether \MODAIC's gains require the full joint design, or whether simpler fixes recover most of the benefit. We evaluate two variants on AVE under the default configuration. \textit{B1} uses load-balanced random selection: agents request updates only from non-saturated peers, but peer choice is uniform and the system lacks \MODAIC's reciprocal modality-decoupled interaction. \textit{B1+B2} further retains delayed peer updates for future aggregation, but still lacks consensus-aware scoring and reciprocal asynchronous interaction. Both variants use the same training budget, model architecture, bandwidth cap, and aggregation cadence as \MODAIC, isolating whether simple load filtering and delayed reuse are sufficient.

\begin{figure}[t]
\centering
\includegraphics[width=.65\linewidth]{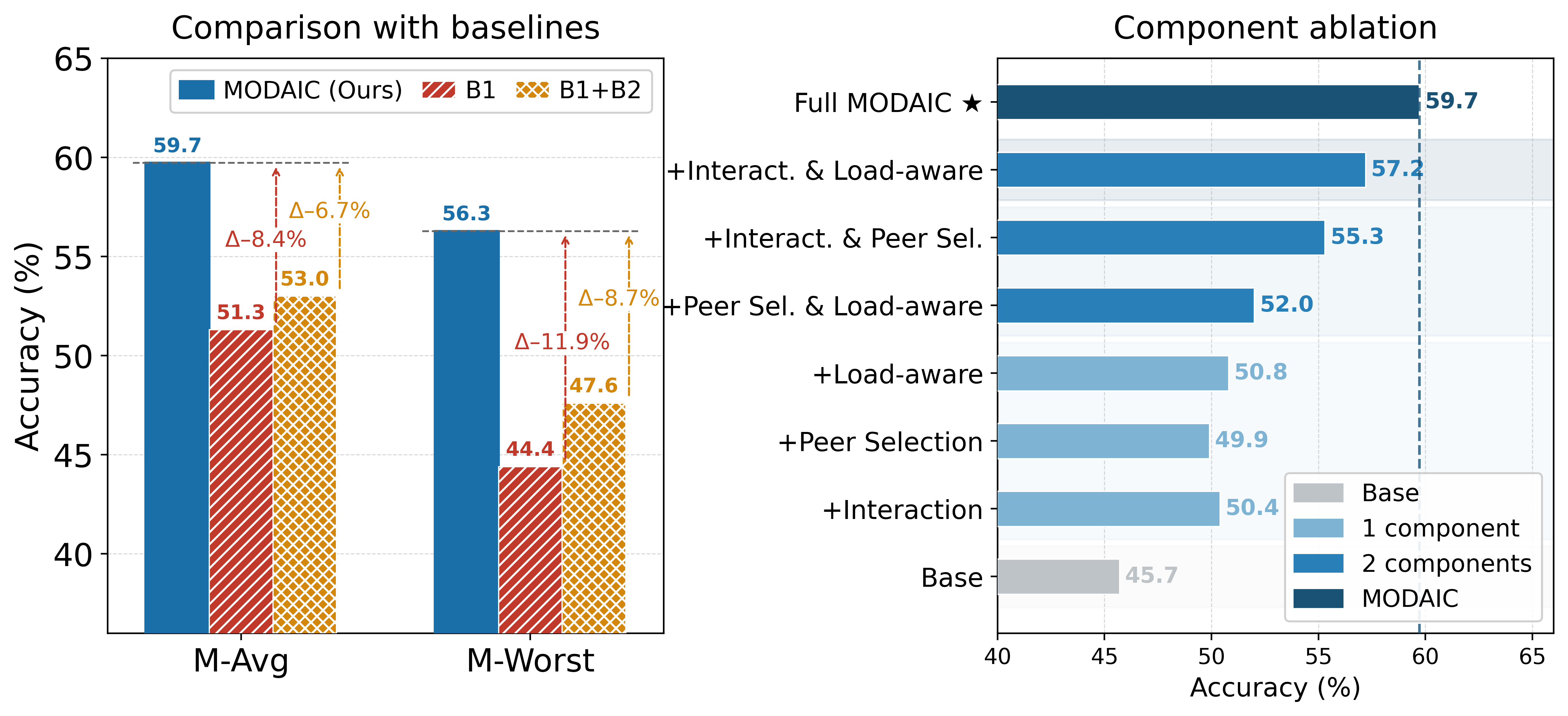}
\vspace{-0.1in}
\caption{\textbf{Simpler variants and component synergy on AVE.} (Left) Load balancing (B1) and delayed aggregation (B1+B2) recover part of the benefit but remain far below \MODAIC, especially on M-Worst. (Right) Component ablations show that local interaction, load-aware fetching, and peer selection address complementary bottlenecks and reinforce one another.}
\label{fig:comp_ablation}
\vspace{-0.1in}
\end{figure}

Figure~\ref{fig:comp_ablation}(left) shows that these fixes help but do not close the gap. B1 trails \MODAIC by 8.4 points in M-Avg and 11.9 points in M-Worst. Adding delayed aggregation improves both metrics, but B1+B2 still trails by 6.7 points in M-Avg and 8.7 points in M-Worst. Thus, the failure is not simply overloaded peers or wasted late transfers: load balancing reduces hot spots and delayed aggregation salvages bytes, but neither preserves reciprocal cross-modal interaction nor selects fresh, pace-compatible peers under deadline pressure.

 \begin{figure}[t]
 \centering
 \includegraphics[width=.65\linewidth]{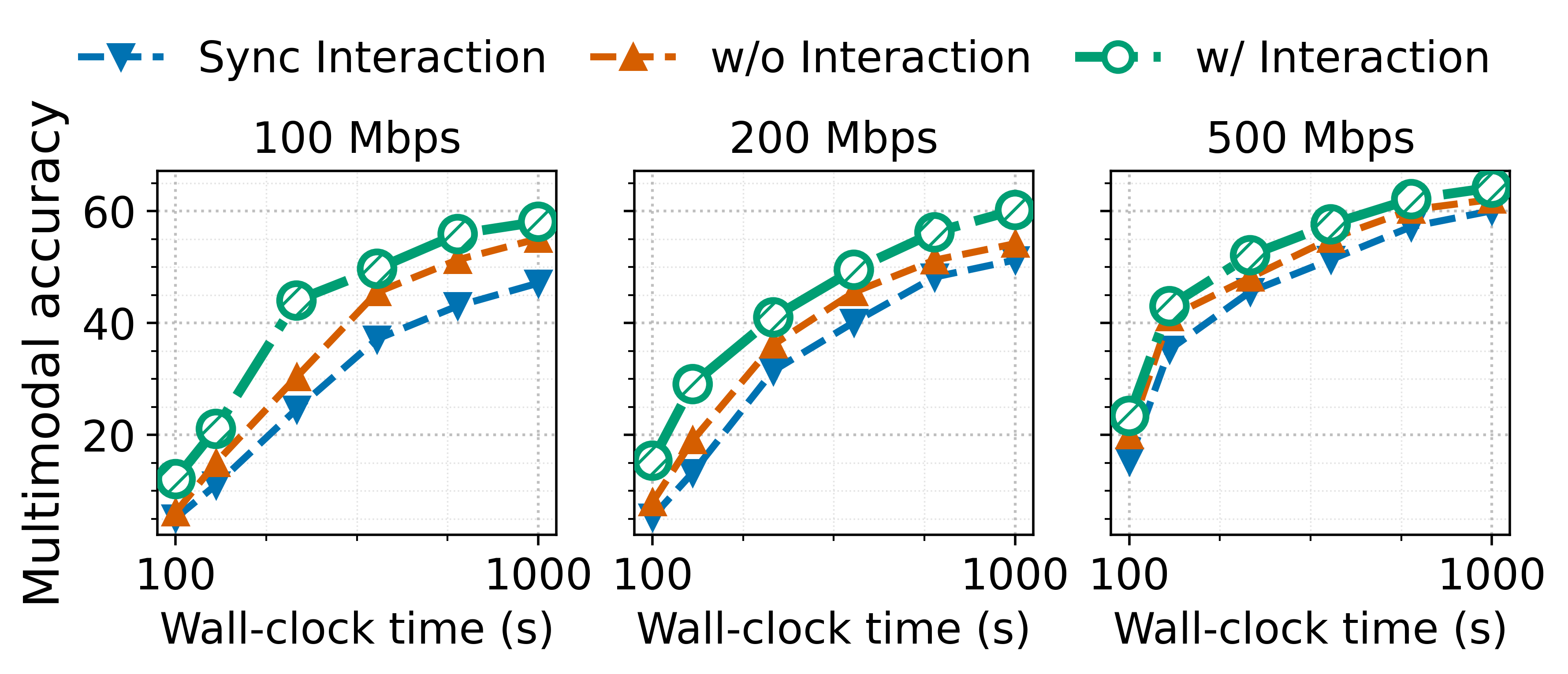}
 \vspace{-0.1in}
 \caption{\textbf{Effect of local interaction block} (multimodal accuracy on AVE).
 Multimodal average accuracy versus wall-clock time at 100, 200, and 500Mbps. \MODAIC's asynchronous interaction outperforms both synchronous interaction and independent modality training over the displayed training horizon.
 }
 \label{fig:interaction_sweep}
 \vspace{-0.1in}
 \end{figure}

\subsection{Component Synergy}\label{sec:component_synergy}

We next examine individual components and their tested combinations. Figure~\ref{fig:comp_ablation}(right) starts from a base asynchronous decentralized system with 45.7\% M-Avg and adds \MODAIC components. Each component improves M-Avg: hierarchical asynchronous collaboration adds 4.7 points, consensus-aware peer selection adds 4.2 points, and load-aware fetching adds 5.1 points. Combinations are stronger than any individual component:
peer selection plus load aware fetching reaches 52.0\% M-Avg, interaction plus peer selection reaches 55.3\% M-Avg, while interaction plus load-aware fetching reaches 57.2\% M-Avg. Full \MODAIC reaches 59.7\%, exceeding these three combinations by 7.7, 4.4 and 2.5 points. The tested combinations therefore show that reciprocal interaction, deadline-resilient exchange, and freshness/fairness-aware peer selection address different bottlenecks and reinforce one another.

\textbf{Local interaction: retaining cross-modal learning without lockstep.}
 We further compare \MODAIC's asynchronous local interaction with two variants: (i) \textit{w/o interaction}, where modalities train independently, and (ii) \textit{sync interaction}, where interaction is enforced only when modalities align in time.
Figure~\ref{fig:interaction_sweep} shows that \MODAIC achieves higher M-Avg than both variants over the plotted training horizon at all three bandwidth budgets. Its advantage over synchronous interaction is larger at 100--200Mbps (where contention and staleness are severe) than at 500Mbps. Synchronous interaction alone does not recover this benefit: in these runs, it also trails training without interaction at the displayed wall-clock checkpoints. These results support the design choice in Section~\ref{subsec:hier_async}: retain cross-modal exchange while allowing modalities to progress independently, matching the goal of Challenge~1.

 \begin{figure}[t]
 \centering
 \includegraphics[width=.65\linewidth]{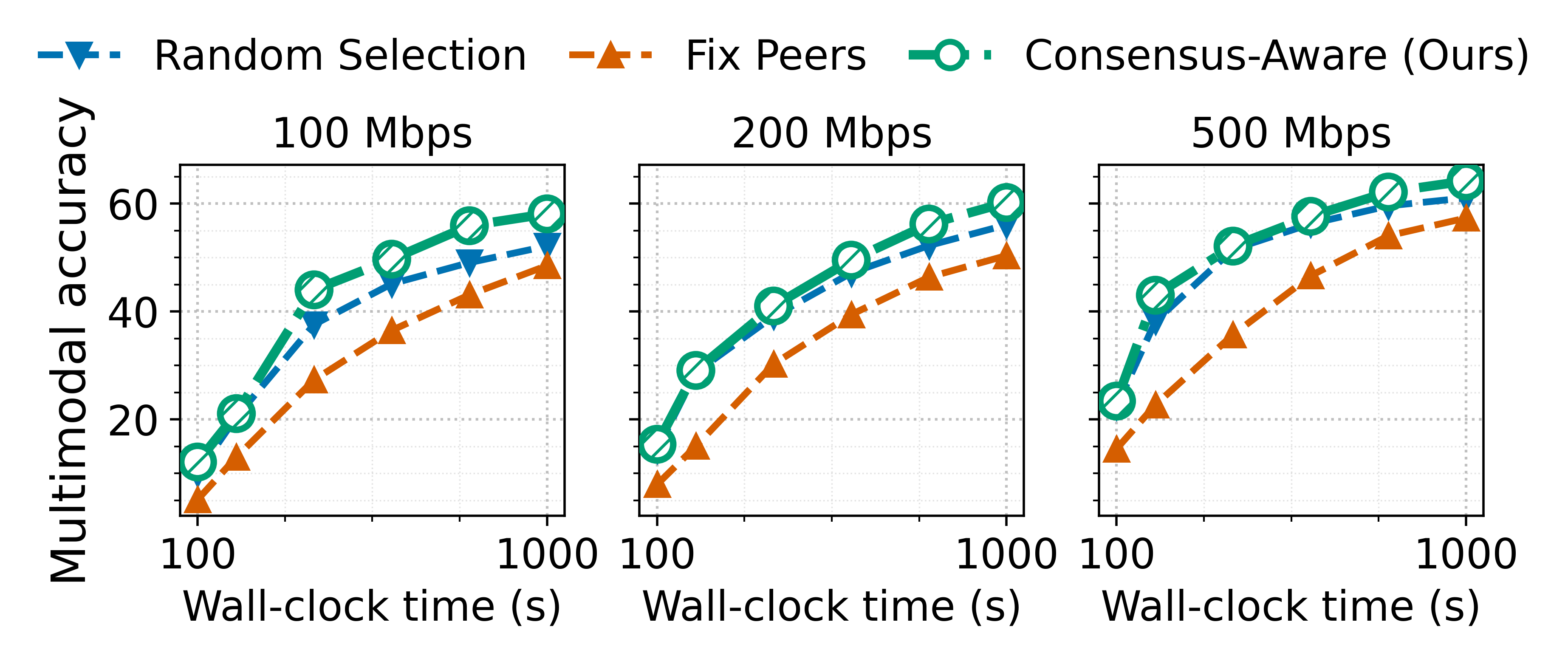}
 \vspace{-0.1in}
 \caption{\textbf{Effect of peer selection} (multimodal accuracy on AVE). Consensus-aware peer selection converges faster than random or fixed neighbors, especially under 100-200~Mbps.}
 \label{fig:selection_sweep}
 \vspace{-0.1in}
 \end{figure}

\textbf{Peer selection: better mixing under bandwidth limits.}
We compare consensus-aware peer selection against \textit{fixed neighbors} (static peer set) and \textit{random selection} (uniform sampling each window). As shown in Figure~\ref{fig:selection_sweep}, consensus-aware peer selection converges faster across bandwidth budgets, with the largest gap at 100-200Mbps. This aligns with Challenge~3: under tight bandwidth, naive selection either overloads popular peers (raising queueing and staleness) or wastes windows on poorly mixed stale updates, whereas \MODAIC uses lightweight log signals to steer requests toward fresher and under-aggregated peers without active probing.

\begin{figure}[t]
\centering
\includegraphics[width=.65\linewidth]{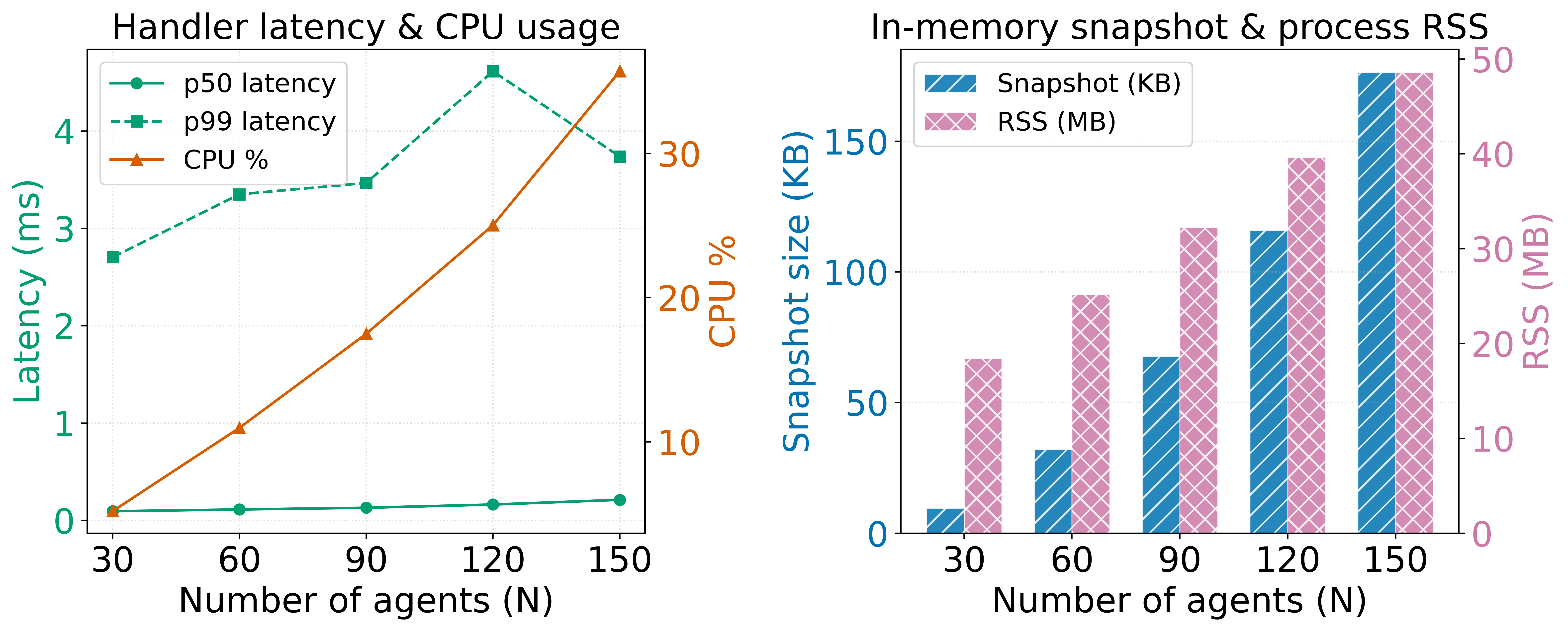}
\vspace{-0.1in}
\caption{\textbf{Tracker overhead.} Under an aggressive 5Hz request rate per agent, handler latency remains below 4.7ms at p99 and CPU remains below 36\% of one core at $N=150$; metadata snapshots remain below 176KB (measurements are based on a tracker with bandwidth of 200Mbps).}
\label{fig:tracker_overhead}
\vspace{-0.1in}
\end{figure}

\subsection{Control-Plane Overhead and Fault Tolerance}\label{sec:tracker_eval}

\MODAIC uses a metadata tracker for admission control and log-guided peer selection, while model states are exchanged directly between peers. We evaluate whether the tracker becomes a practical bottleneck using the same single-process Python tracker as in the training experiments. In a stress test, each agent issues tracker requests at 5Hz, higher than in our training runs, while we vary the number of agents from 30 to 150. Figure~\ref{fig:tracker_overhead} shows that the tracker handles the offered load with p50 latency below 0.22ms and p99 latency below 4.7ms. CPU usage grows to 35.7\% of one core at 150 agents, while the in-memory snapshot grows only from 9.6 to 176KB and process RSS from 18MB to 49MB. Thus, at the evaluated scale and under an aggressive request rate, the tracker is not the limiting resource and its state is small enough to replicate or shard by modality.

We also evaluate tracker fault tolerance. When the tracker is periodically unavailable to a fraction of agents, agents continue using their last admissible-peer snapshot and cached updates. Across 70--90\% accessibility, \MODAIC follows a similar convergence trend and remains competitive. Injecting 0.1--0.5s tracker delays increases time-to-convergence by at most 50s and reduces final accuracy by at most 0.8\%, indicating that tracker quality affects peer freshness but does not block local training. Full results are in Appendix~\ref{sec:tracker_eval_app}.

\subsection{Sensitivity and Robustness}\label{sec:sensitivity}

We test whether \MODAIC remains robust under bandwidth limits and WAN-like dynamics.

\begin{figure}
    \centering
    \includegraphics[width=.65\linewidth]{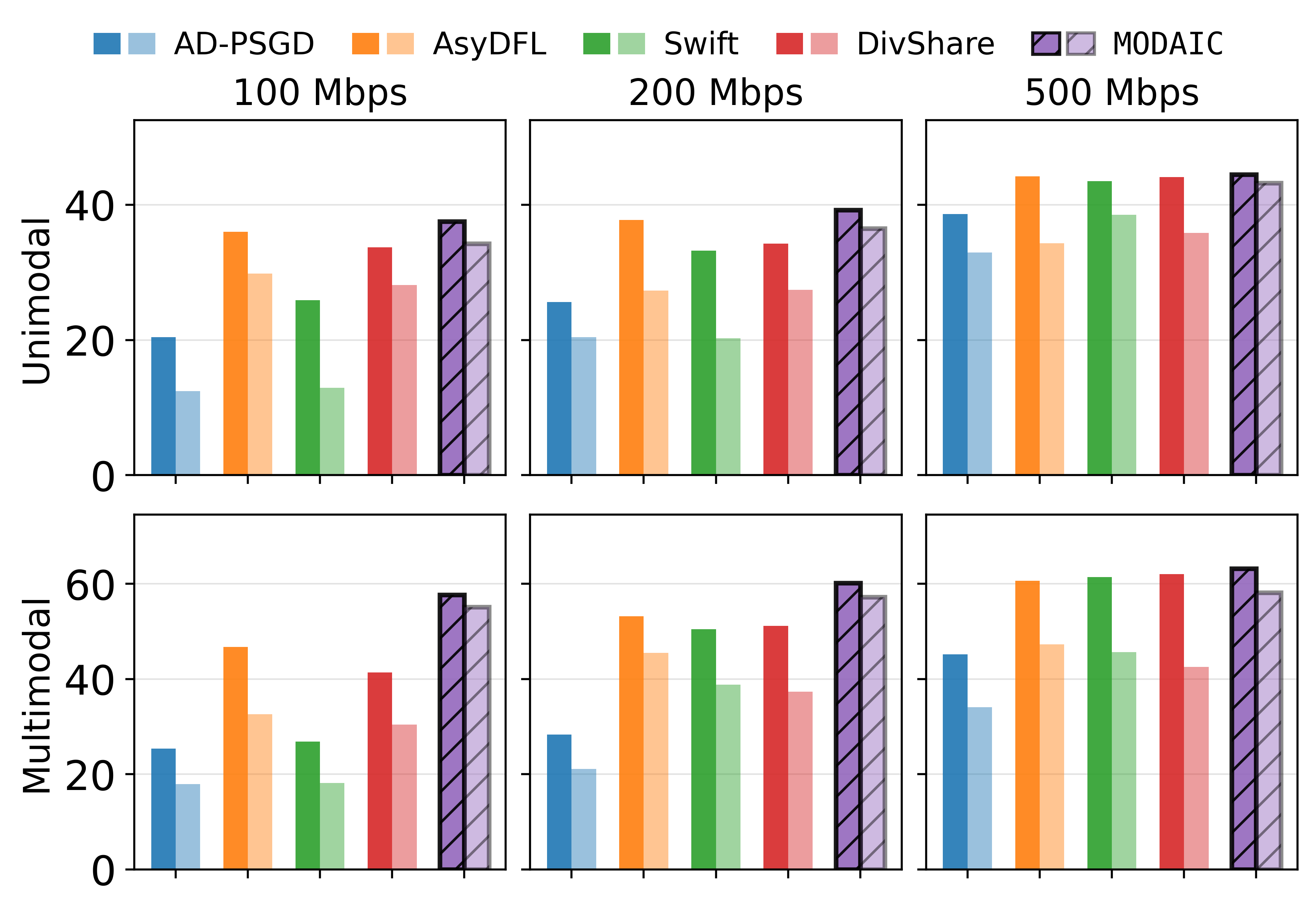}
   \vspace{-0.1in}
        \caption{\textbf{Bandwidth sweep} (Dir. $\alpha=0.5$). Columns vary per-agent bandwidth; rows show unimodal (top) and multimodal (bottom) accuracy. Solid bars are averages across agents, and light/hatched bars are worst-agent accuracy.}
    \label{fig:bandwidth_sweep}
       \vspace{-0.1in}
\end{figure}

\textbf{Bandwidth sweep.}
Figure~\ref{fig:bandwidth_sweep} varies the per-agent bandwidth budget under the default non-IID setting (Dir. $\alpha=0.5$). Across 100-500Mbps, \MODAIC consistently achieves higher average accuracy while substantially improving worst-agent accuracy, yielding a smaller Avg-Worst gap for both unimodal and multimodal agents. The advantage is largest under tight bandwidth (100-200Mbps), where many-to-one contention and missed aggregation deadlines dominate. \MODAIC mitigates these effects via contention-aware pulls with admission control and by reusing late arrivals through caching, so slow or overloaded agents are not systematically under-mixed.

\begin{figure}[t]
\centering
\includegraphics[width=0.65\linewidth]{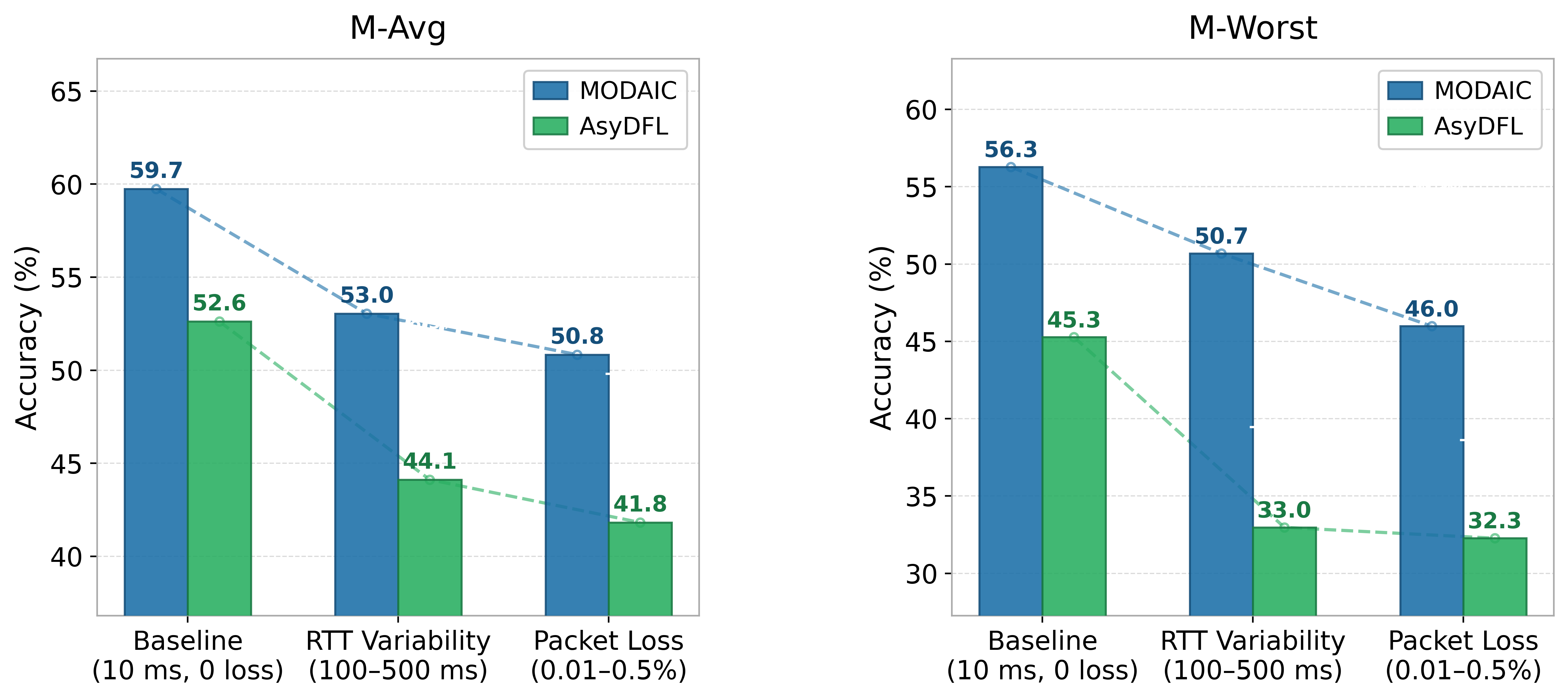}
\vspace{-0.1in}
\caption{\textbf{WAN stress conditions on AVE.} We compare \MODAIC with AsyDFL, the strongest baseline, under RTT variability and packet loss. Performance drops are measured relative to the 10ms, zero-loss baseline.}
\label{fig:wan_stress}
\vspace{-0.1in}
\end{figure}

\textbf{WAN-style RTT variation and packet loss.}
To test robustness beyond fixed-bandwidth emulation, we inject two WAN stress conditions: (i) RTT variability sampled uniformly between 100ms and 500ms across node pairs, and (ii) packet loss rates from 0.01\% to 0.5\%. Figure~\ref{fig:wan_stress} compares \MODAIC with AsyDFL, the strongest baseline in the default AVE setting. Under RTT variability, \MODAIC's M-Avg drops by 6.7 points and M-Worst by 5.6 points, compared to 8.5 and 12.3 points for AsyDFL. Under packet loss, \MODAIC drops by 8.9/10.3 points (M-Avg/M-Worst), compared to 10.8/13.0 points for AsyDFL. The worst-agent gap widens from 11.0 points at baseline to 17.7 points under RTT variability and 13.7 points under packet loss, confirming that \MODAIC's modality decoupling and load-aware scheduling are particularly beneficial for stragglers under realistic network dynamics.

\textbf{Stragglers and system scale.}
Appendix~\ref{sec:sensitivity_app} separately tests communication stragglers by reducing selected agents' bandwidth fourfold, and computation stragglers by slowing local training steps fivefold through added sleep. \MODAIC retains the highest U-Avg and M-Avg among the compared methods at the tested straggler fractions. On KU-HAR, it also retains the highest group averages as the population grows from 30 to 90 agents at 1000Mbps. Each agent receives one of 90 predefined data shards, so the larger deployments also contain more training data. This experiment tests operation with larger participating populations rather than isolating scaling at a fixed total dataset size.

\subsection{Wide-Area Cloud Deployment}\label{sec:cloud_exp}

\textbf{Cloud testbed.}
We deploy a cross-region AWS testbed with three EC2 instances in Virginia (US), London (UK), and Singapore. Each host runs multiple \MODAIC agents as Docker containers (CPU-only) and communicates over an encrypted P2P overlay using TCP connections. We enforce per-container bandwidth caps using Linux traffic control to emulate heterogeneous wide-area links. All other settings follow the local testbed unless noted. \textit{This cloud setup is not intended to demonstrate large-scale throughput; instead, it validates that \MODAIC's coordination plane and pull-based runtime operate reliably over real WAN latency/jitter and container networking, and that its gains persist when computation dominates.} In our deployment, inter-region RTTs range from 158 to 226ms across region pairs.

\begin{figure}
    \centering
    \includegraphics[width=.65\linewidth]{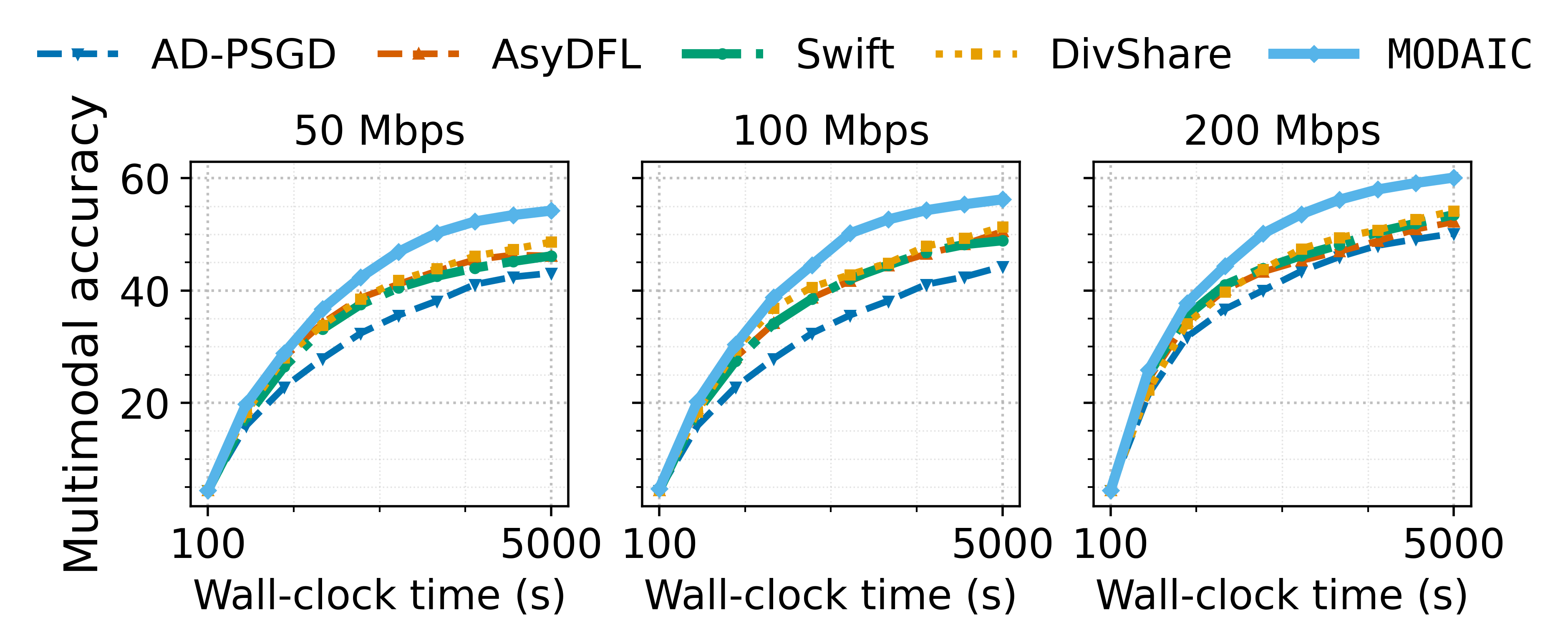}
    \vspace{-0.1in}
     \caption{\textbf{Cloud learning curves on AVE} (CPU-only). Multimodal test accuracy vs. wall-clock time under different bandwidth caps.}
    \label{fig:learning_curve_aws}
    \vspace{-0.1in}
\end{figure}

\begin{figure}
    \centering
    \includegraphics[width=.65\linewidth]{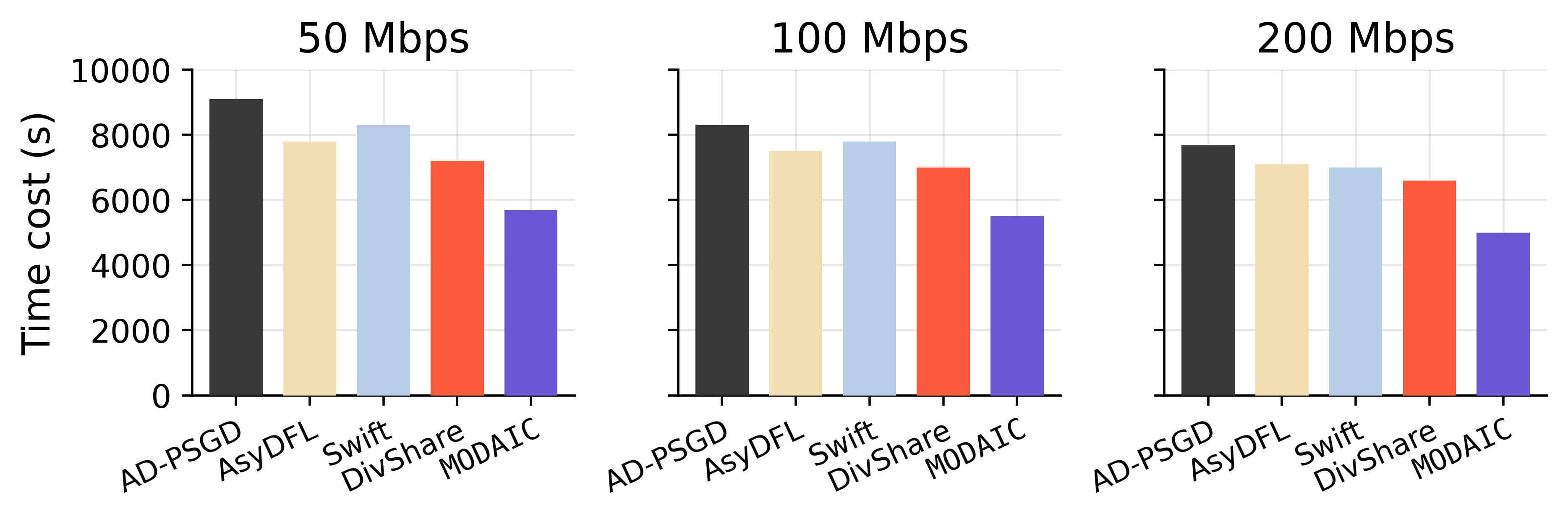}
    \vspace{-0.1in}
      \caption{\textbf{Time-to-accuracy on the cloud testbed.} Wall-clock time to reach a fixed multimodal accuracy target (M-Avg), under different bandwidth caps.}
    \label{fig:time_cost_aws}
    \vspace{-0.1in}
\end{figure}

\textbf{Learning progress and time-to-accuracy.}
Figure~\ref{fig:learning_curve_aws} shows that \MODAIC converges faster than all baselines across bandwidth settings. Figure~\ref{fig:time_cost_aws} further shows that, to reach the same multimodal accuracy target, \MODAIC reduces wall-clock training time by up to 30\%. These results indicate that \MODAIC improves optimization efficiency even when computation, rather than communication, is the dominant bottleneck, and that its coordination plane and fetching policy remain effective under real wide-area deployment dynamics. Appendix~\ref{sec:cloud_exp_app} reports the full cloud bandwidth sweep. These results extend the local-testbed findings to CPU-based training over real cross-region links, supporting the practicality of \MODAIC's coordination and peer-to-peer update exchange beyond the controlled GPU setting.

\section{Related Work}\label{sec:related}

\textbf{Decentralized learning.}
Compared with server-based collaboration~\cite{MLSYS2024_bbd7d8bd,tang2025fedprophet,he2025hourglass,ching2024totoro,liu2025adaptive}, decentralized learning exchanges updates directly among peers, avoiding centralized model aggregation and supporting geographically distributed deployments~\cite{liao2023adaptive,zhang2023poster}. Prior work develops asynchronous protocols and convergence analyses under heterogeneous progress and update delays~\cite{ad-psgd,pmlr-v238-even24a,bornstein2023swift,biswas2025boosting}. Other work studies communication topology~\cite{nedic2018network,dynamic-topo,epidemic-topo} and resource-aware neighbor selection~\cite{asydfl}. Related FL work optimizes participant selection~\cite{lai2021oort,pmlr-v151-jee-cho22a} and resource allocation across learning jobs~\cite{liu2025venn}. \MODAIC targets modality-structured imbalance: agents essential for cross-modal knowledge transfer also carry heavier computation and communication workloads across multiple modality overlays.

\textbf{Multimodal learning systems.}
Multimodal learning combines complementary signals (e.g., audio and video) for robustness and accuracy~\cite{cui2024survey,moin2023emotion,avedataset,xia2022cross,zhang2020pdlens}. Fusion and cross-modal alignment objectives couple modality-specific encoders through cross-modal representations~\cite{radford2021learning,jia2021scaling,xu2023mesen}. Recent systems improve distributed training throughput through heterogeneity-aware parallelism, placement, and scheduling~\cite{distmm24,zhang2025disttrain}. These systems optimize distributed execution of multimodal models; \MODAIC addresses decentralized agents with different modality subsets, where independent local progress must preserve reciprocal cross-modal interaction.

\textbf{Multimodal federated/decentralized learning under heterogeneity.}
Under \emph{modality heterogeneity}, agents hold different modality subsets, introducing learning imbalance and uneven resource demands. Multimodal FL frameworks address this heterogeneity through server-mediated training and aggregation~\cite{wang2024fedhkd,yang2024MCARN}. Harmony~\cite{ouyang2023harmony} disentangles encoder and fusion training and coordinates modalities with different training delays. Recent multimodal FL work also addresses incomplete modalities through server-coordinated learning~\cite{liu2025fedmobile,wang2025multimodal}. In decentralized settings, DMML-KD~\cite{kd-dfl} combines knowledge distillation with modality-dependent local iteration control to balance training. \textit{\MODAIC differs by jointly coordinating reciprocal cross-modal interaction and wait-free update exchange under shared bandwidth.} It combines load-aware admission control, late-arrival reuse, and metadata-guided peer selection without active probing. Model exchange remains peer-to-peer, while a lightweight tracker handles coordination metadata.

\section{Conclusions}\label{sec:conclusion}
We presented \MODAIC, a wait-free runtime that coordinates reciprocal cross-modal interaction, load-aware update exchange, and peer selection without active probing under modality heterogeneity. Experiments on AVE and KU-HAR across local and cross-region testbeds show higher average and multimodal worst-agent accuracy and shorter time-to-accuracy. Ablations show that load balancing and delayed aggregation leave a worst-agent gap, while tracker experiments demonstrate modest processing overhead and tolerance to partial unavailability.

\clearpage
\bibliographystyle{ACM-Reference-Format}
\bibliography{references}

\appendix

\section{Additional Motivation Results}\label{sec:motivation-app}

\subsection{Cross-modal Interaction Ablation}\label{app:interaction_ablation}

\noindent\textbf{Setup.}
We compare three interaction designs while keeping the same model architecture, data split, and aggregation schedule. The only difference is how cross-modal features are exchanged and used to form the interaction/alignment loss: (i) no interaction, (ii) uni-directional interaction, and (iii) bi-directional (reciprocal) interaction.

\begin{figure}[h]
    \centering
    \includegraphics[width=0.65\linewidth]{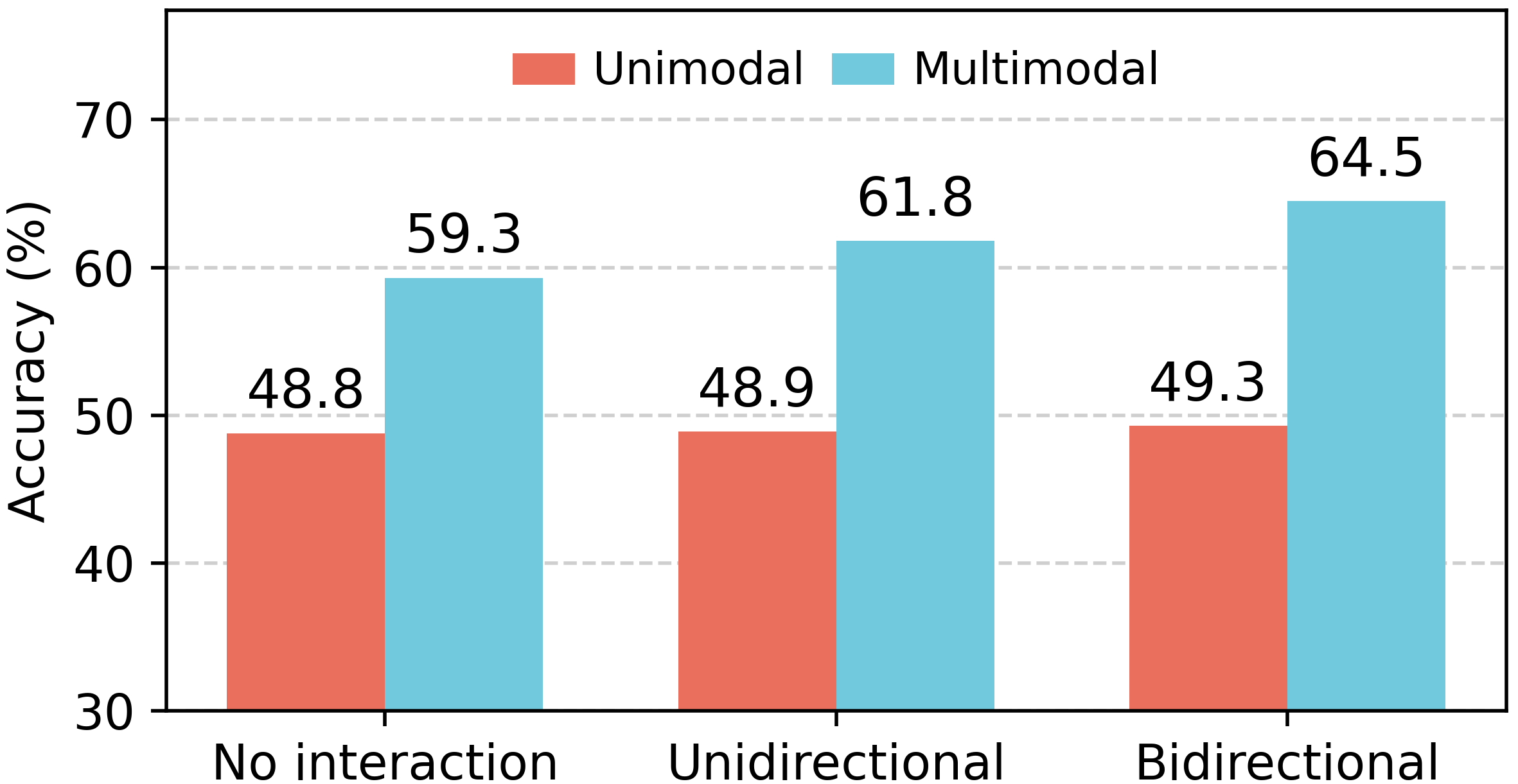}
    \caption{Cross-modal interaction variants. \emph{No interaction} trains each modality independently; \emph{uni-directional} aligns one modality to the other; \emph{bi-directional} enables reciprocal alignment where each modality both produces and consumes cross-modal features. To isolate interaction effects, we use the same (synchronous) aggregation schedule across variants.}
    \label{fig:interaction_plot_app}
\end{figure}

\noindent\textbf{Result and implication.}
Figure~\ref{fig:interaction_plot_app} shows that removing interaction degrades multimodal accuracy, and uni-directional alignment recovers only part of the gap. Bi-directional interaction consistently performs best, indicating that reciprocal exchange is important for stable cross-modal alignment in our setting. This supports the design goal in Challenge~1: rather than eliminating cross-modal coupling, the system must preserve \emph{reciprocal} interaction while avoiding the synchronization-induced stalls that make multimodal agents straggle.

\subsection{Communication delay breakdown and implications}\label{app:comm_breakdown}
Figure~\ref{fig:modality_comm}(right) shows that the dominant component of per-exchange completion time is wait/serialization (enqueue-to-send), while the bandwidth-limited transfer time is small. This indicates that under modality heterogeneity, performance is often limited by contention and multiplexing at hot-spot peers rather than raw link capacity. Consequently, synchronous protocols amplify multimodal agents as bottlenecks, while wait-free protocols under-sample them due to missed windows.

\subsection{Additional discussion: scheduling limits under heterogeneity}
\label{app:sched_limits}
Under modality heterogeneity, the feasible set of timely peers changes over time due to compute skew and multi-overlay contention. Random selection wastes limited exchanges on late/stale updates, while fixed topologies cannot react to transient hot spots and stragglers. Although probing peers for queue occupancy or pace could inform better selection, active probing adds control traffic and RTTs and can itself create contention, motivating lightweight shared progress signals and late-update reuse.

\section{Additional Details on \MODAIC Design}\label{app:modaic}

\subsection{Local asynchronous interaction details}\label{app:async_interaction}

\noindent\textbf{Credit-and-buffer procedure.}
Algorithm~\ref{alg:model_interact} gives the exact contribute-before-consume logic used in Section~\ref{subsec:hier_async}. Each modality $m$ maintains a single-slot buffer $\textsc{Buf}[m]$ storing $(x_o,y,h_m)$ from its most recent paired sample and a boolean flag $\textsc{Credit}[m]$. If $\textsc{Credit}[m]$ is true and $\textsc{Buf}[o]\neq\varnothing$, modality $m$ consumes $(x_m,y,h_o)$ from $\textsc{Buf}[o]$ (matched by construction), computes $h_m$, and updates using $\mathrm{SL}(h_m,y)+\mathrm{IL}(h_m,h_o)$. Otherwise it performs a supervised step on the next paired batch and enqueues $(x_o,y,h_m)$ into $\textsc{Buf}[m]$ if empty. The single-slot buffer bounds memory and prevents long queues of stale features under heavy skew.

\begin{algorithm}[t]
\caption{Asynchronous cross-modal interaction}
\label{alg:model_interact}
\begin{algorithmic}[1]
\RequireH{Current modality $m$, the other modality $o$, two data streams (dataloaders) $\mathcal{D}_o, \mathcal{D}_m$, feature buffer $\textsc{Buf}[m], \textsc{Buf}[o]$, credit $\textsc{Credit}[m]$, supervised loss $\mathrm{SL}(\cdot)$, interaction loss $\mathrm{IL}(\cdot)$}

\Procedure{TrainStep}{$m$}
\If{$\textsc{Credit}[m]$ \textbf{and} $\textsc{Buf}[o]\neq\varnothing$}
    \State $(x_m, y, h_o) \gets \textsc{Buf}[o]$
    \State $h_m \gets f_{\theta_m}(x_m;\, m)$
    \State $\textsc{Buf}[o] \gets \varnothing$
    \State $\textsc{Credit}[m] \gets \texttt{false}$
    \State $\mathcal{L} \gets \mathrm{SL}(h_m, y) + \mathrm{IL}(h_m, h_o)$
\Else
    \State $(x_m, x_o, y) \gets \textsc{NextBatch}(\mathcal{D}_m)$
    \State $h_m \gets f_{\theta_m}(x_m;\, m)$
    \State $\mathcal{L} \gets \mathrm{SL}(h_m, y)$
    \If{$\textsc{Buf}[m]=\varnothing$}
        \State $\textsc{Buf}[m] \gets (x_o, y, h_m)$
        \State $\textsc{Credit}[m] \gets \texttt{true}$
    \EndIf
\EndIf

\State update $\theta_m$ using $\nabla_{\theta_m} \mathcal{L}$
\EndProcedure
\end{algorithmic}
\end{algorithm}

\subsection{Cache management details}\label{app:cache_details}
Each agent maintains a per-modality receive cache $\mathcal{B}_i^m$ of recently completed peer updates (or deltas) for modality $m$. At each aggregation deadline, the agent selects up to $K$ cached updates according to the peer-selection policy (Section~\ref{subsec:peer_selection_design}); unselected entries remain eligible for future windows. Upon aggregation, selected entries are removed from $\mathcal{B}_i^m$ and recorded in an aggregated set $\mathcal{A}_i^m$. To bound memory, we evict cache entries by age and cap each $\mathcal{B}_i^m$ to $O(K)$ items (we use $4K$ in our implementation); we likewise bound $\mathcal{A}_i^m$ with a sliding window over recent versions/steps so bookkeeping stays lightweight.

\subsection{Peer selection pseudocode}\label{app:peer_selection}
Algorithm~\ref{alg:consensus_peer_select} gives the full selection procedure, including construction of $\mathcal{C}^m_i(t)$ from the coordination log, fallback to the on-device cache pool $\mathcal{A}^m_i(t)$ when $|\mathcal{B}^m_i(t)|<K$, and optional loss-difference reweighting with $\beta_w$.

\begin{algorithm}[t]
\caption{Consensus-aware peer selection (per-modality)}
\label{alg:consensus_peer_select}
\begin{algorithmic}[1]
\RequireH{Agent $i$, modality $m$ at deadline $t$, log $\mathcal{J}$ entries, last-round aggregated IDs $\mathcal{H}^m_i$, counters $\mathrm{count}^m_{ij}$, base weights $W^m_{ij}$, aggregation budget $K$, hyperparams $\alpha\in[0,1]$, $\beta>0$, $\beta_w>0$}

\State $\mathcal{E}^m_i \gets \{i\}\cup \mathcal{H}^m_i$
\State $\mathcal{C}^m_i \gets \textsc{LatestUnaggregated}(\mathcal{J}, t, \mathcal{E}^m_i)$
\State $\mathcal{A}^m_i \gets \textsc{LatestFallback}(\mathcal{J}, t, \{i\}\cup \mathcal{C}^m_i)$

\ForAll{$j \in \mathcal{C}^m_i$}
\Comment{Compute selection scores}
  \State $s^m_{ij} \gets (1-\alpha)\cdot\frac{1}{1+\mathrm{count}^m_{ij}}
  \;+\;\alpha\cdot\exp\!\big(-|\ell^m_i-\ell^m_j|/\beta\big)$
\EndFor
\State $S^m_i \gets \textsc{Sample}(\mathcal{C}^m_i,\; \min(K,|\mathcal{C}^m_i|),\; \propto s^m_{ij})$
\If{$|\mathcal{B}^m_i| < K$} \Comment{After local training}
  \State $\mathcal{K}^m_i \gets \mathcal{B}^m_i \cup \textsc{Sample}(\mathcal{A}^m_i,\; K-|\mathcal{B}^m_i|,\; \propto s^m_{ij})$
\EndIf
\ForAll{$j \in \mathcal{K}_i$}
  \State $\tilde{w}^m_{ij} \gets W^m_{ij}\cdot \exp\!\big(-|\ell^m_i-\ell^m_j|/\beta_w\big)$
  \Comment{Reweighting}
\EndFor
\State $w^m_{ij} \gets \tilde{w}^m_{ij}\Big/\sum_{u\in \mathcal{K}^m_i\cup\{i\}} \tilde{w}^m_{iu}$
\State $\theta_m \gets \sum_{j\in \mathcal{K}^m_i\cup\{i\}} w^m_{ij}\,\theta_j$ \Comment{Model aggregation}
\State $\mathcal{B}^m_i \gets \mathcal{B}^m_i \setminus \mathcal{K}^m_i$
\State $\mathcal{A}^m_i \gets \mathcal{A}^m_i \cup \mathcal{K}^m_i$
\State \Return $(S^m_i,\{w^m_{ij}\})$

\end{algorithmic}
\end{algorithm}

\paragraph{Fallback sampling distribution.}
When sampling from $\mathcal{A}^m_i(t)$, we reuse Eq.~\ref{eq:score} to define $s^m_{ij}$ for each cached update $j\in\mathcal{A}^m_i(t)$ and sample without replacement with probability proportional to $s_{ij}$, i.e.,
$p^m_{ij} = s_{ij} / \sum_{u\in\mathcal{A}^m_i(t)} s_{iu}$.

\section{Additional Implementation Details}\label{app:impl_details}

\noindent\textbf{Baseline Fairness.}
For asynchronous decentralized baselines (AD-PSGD~\cite{ad-psgd}, AsyDFL~\cite{asydfl}, Swift~\cite{bornstein2023swift}, and DivShare~\cite{biswas2025boosting}), we follow the original push/pull protocols. To avoid confounding effects from topology and pairwise contention, we run them on a fully connected overlay with dedicated per-pair links (i.e., no link-level contention). For AD-PSGD, we use its standard random bipartite active--passive graph, so active agents can always communicate with passive agents.

For synchronous baselines, we use the default settings in the original papers. FedLay~\cite{hua2024towards} operates on per-modality overlays; in each round, an agent selects $K$ neighbors within each overlay to exchange the corresponding modality parameters. For DMML-KD~\cite{kd-dfl}, all agents participate in a single fully connected overlay, and the multimodal fusion module follows the method design. Harmony~\cite{ouyang2023harmony} is originally centralized; we adapt it by replacing the server with a lightweight tracker for stage coordination, while updates are exchanged via random gossip within per-modality overlays. In Harmony’s second stage, unimodal agents communicate only with same-modality peers, and multimodal agents communicate only with multimodal peers.

Unless otherwise specified, all baselines (except DMML-KD) use the same modality-grouping rule as \MODAIC: agents sharing a modality join the same overlay, and multimodal training uses the same aligned fusion objective (contrastive feature alignment~\cite{radford2021learning}).

\noindent\textbf{Training Setting.} All trainable modules, including the feature extractors and classifiers, are optimized with SGD (learning rate $10^{-3}$, momentum 0.9, weight decay $5\times10^{-4}$) using a batch size of 32 in all experiments. In each communication round, every agent performs a single local SGD update step.

\section{Additional Experimental Results}\label{app:extra_results}
\subsection{Additional System Cost Results}

\begin{figure}
    \centering
    \includegraphics[width=0.65\linewidth]{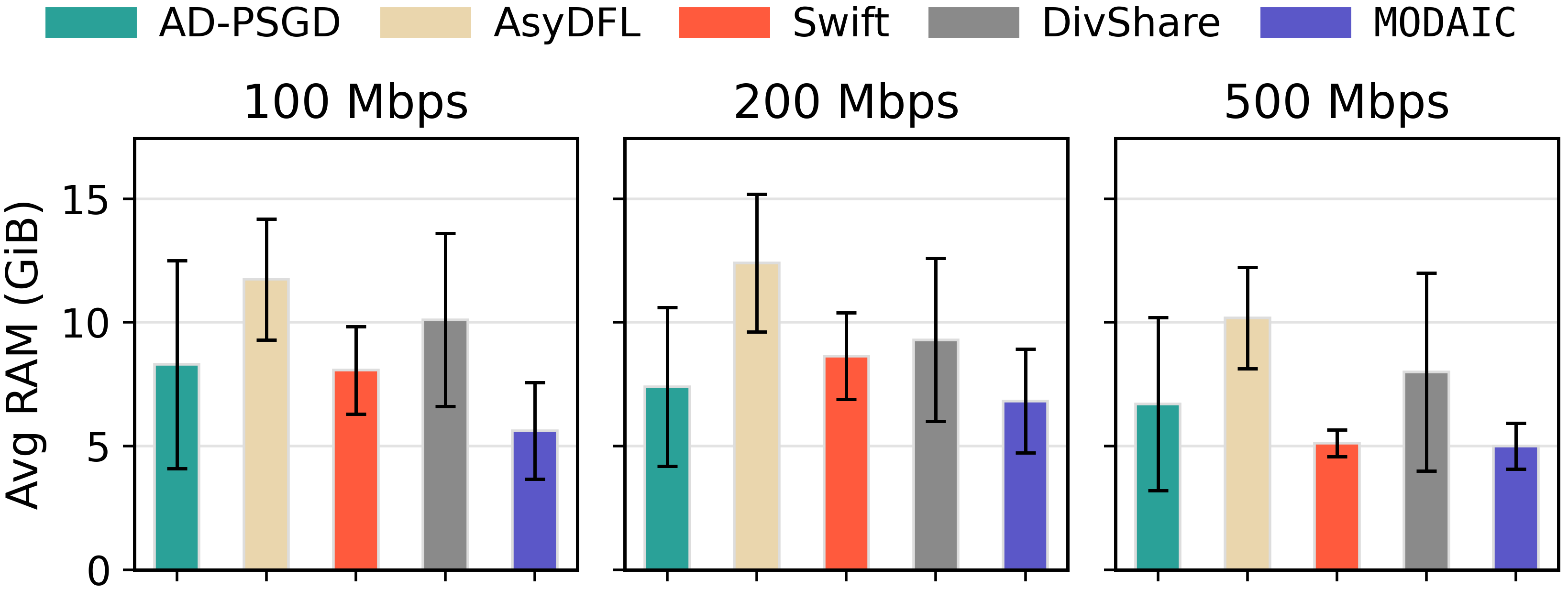}
    \caption{The peak memory usage of multimodal agents of different methods.}
    \label{fig:mem_comp}
\end{figure}

\noindent\textbf{Memory usage comparison.} Following the default setting in Section~\ref{sec:main_results}, we measure the RAM consumption of multimodal agents in Figure~\ref{fig:mem_comp}. The memory footprint is dominated by three components: cached peer updates, the upload buffer, and the download buffer. Compared with the baselines, our method reduces memory usage by up to $50\%$ (i.e., about $2\times$). At higher bandwidth (500Mbps), it still maintains comparably low memory usage, nearly matching the most memory-efficient baseline. These results demonstrate the memory efficiency of our approach.

\subsection{Additional Control-Plane Overhead and Fault Tolerance}\label{sec:tracker_eval_app}

\begin{figure}[t]
\centering
\includegraphics[width=0.65\linewidth]{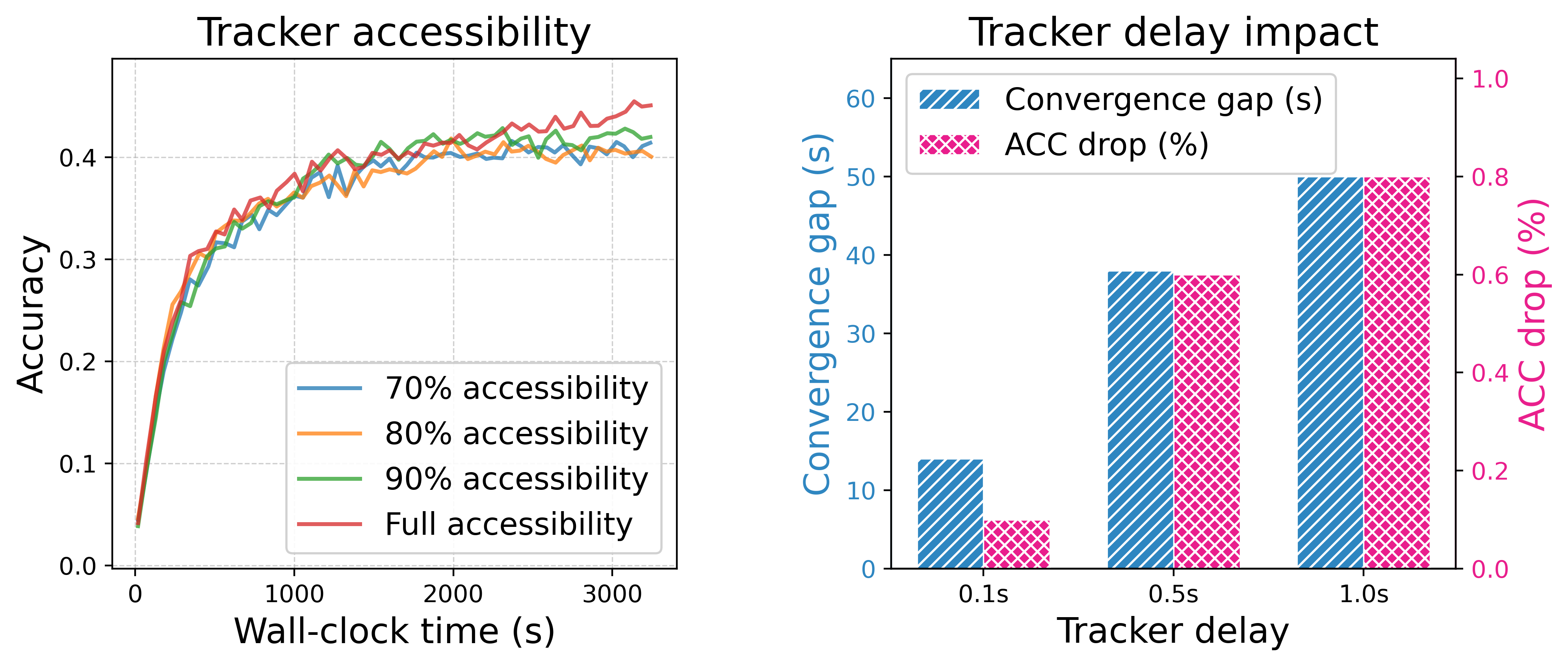}
\vspace{-0.1in}
\caption{\textbf{Tracker fault tolerance.} (Left) Training under partial tracker accessibility. Agents without tracker access use the last admissible snapshot and cached updates. (Right) Injected tracker delay increases convergence time and accuracy drop gradually rather than causing failure.}
\label{fig:tracker_access}
\vspace{-0.15in}
\end{figure}

We also evaluate fault tolerance. At each 10-second window, we randomly make the tracker unreachable to a fraction of agents; those agents continue using their last admissible-peer snapshot and cached updates. Figure~\ref{fig:tracker_access}(left) shows that all accessibility levels follow similar early convergence, and even 70\% accessibility converges to a competitive final accuracy. Figure~\ref{fig:tracker_access}(right) injects tracker response delays of 0.1-0.5s. \MODAIC degrades gracefully: time-to-convergence increases by up to 50s and final accuracy drops by at most 0.8\%. These results support the design goal that tracker quality affects peer freshness, but tracker delay or temporary unavailability does not block local training.

\subsection{Additional Sensitivity and Robustness}\label{sec:sensitivity_app}

\begin{figure}
    \centering
    \includegraphics[width=.65\linewidth]{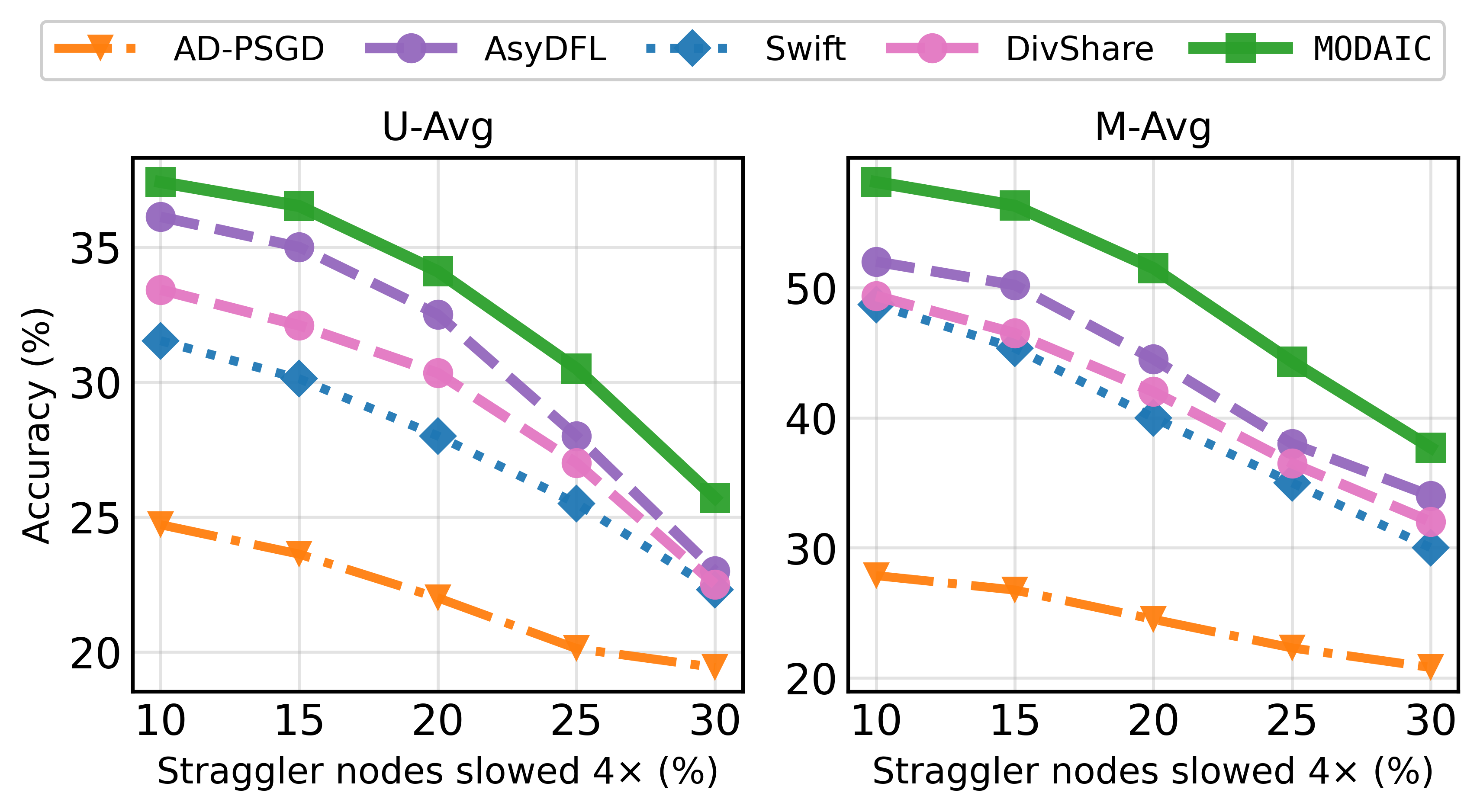}
    \vspace{-0.1in}
     \caption{\textbf{Communication stragglers:} a fraction of agents have $4\times$ lower bandwidth. }
    \label{fig:comm_straggler}
    \vspace{-0.15in}
\end{figure}

\begin{figure}
    \centering
    \includegraphics[width=.65\linewidth]{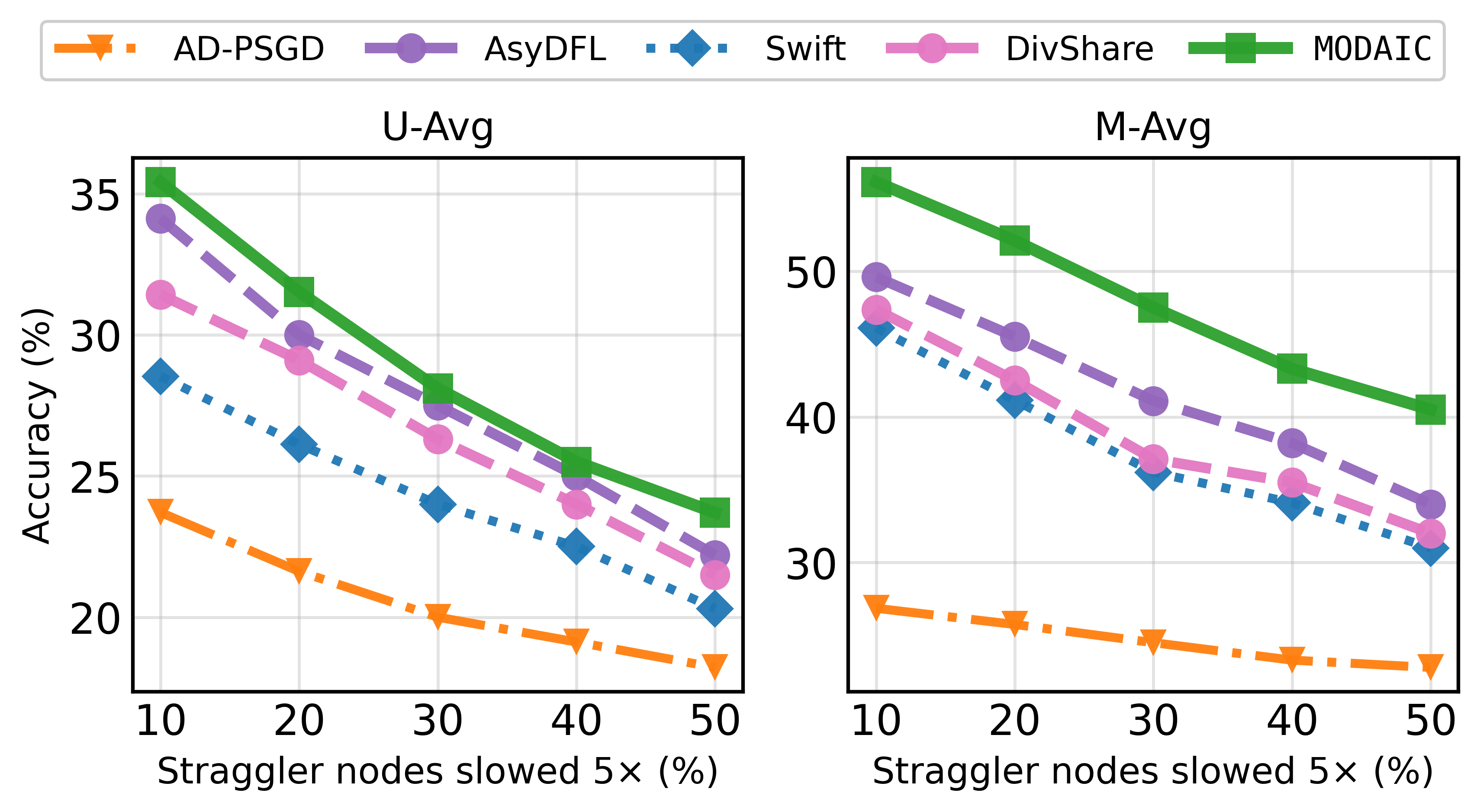}
    \vspace{-0.1in}
    \caption{\textbf{Computation stragglers:} a fraction of agents run $5\times$ slower local steps (emulated by added sleep), with bandwidth fixed at 200Mbps.}
    \label{fig:comp_straggler}
    \vspace{-0.15in}
\end{figure}

\noindent\textbf{Straggler analysis.}
Under the default 200Mbps setting, we evaluate robustness to \emph{communication} stragglers by assigning a fraction of agents $4\times$ lower bandwidth. Figure~\ref{fig:comm_straggler} shows that accuracy degrades for all methods as the straggler fraction increases, but \MODAIC preserves a stable margin, especially on multimodal metrics.  Communication stragglers are particularly damaging because slow links create in-flight backlog: more transfers complete after aggregation deadlines, fewer updates are actually applied on time, and effective staleness increases, which disproportionately hurts the worst agents. \MODAIC mitigates this via request-driven fetching with lightweight admission control and by caching late arrivals for reuse instead of wasting transmitted bytes.

We next evaluate \emph{computation} stragglers by keeping bandwidth fixed and making a fraction of agents $5\times$ slower in local training (emulated by extra sleep in the training thread). As shown in Figure~\ref{fig:comp_straggler}, the degradation pattern differs: slow agents produce updates less frequently, but links are not congested, so when exchanges occur they can still deliver relatively fresh updates. Even in this regime, \MODAIC consistently improves accuracy, indicating that hierarchical modality-decoupled asynchrony reduces compute-induced staleness while preserving bidirectional interaction, and consensus-aware peer selection continues to favor fresher, pace-similar peers as progress diverges.

\begin{figure}[t]
    \centering
    \includegraphics[width=.65\linewidth]{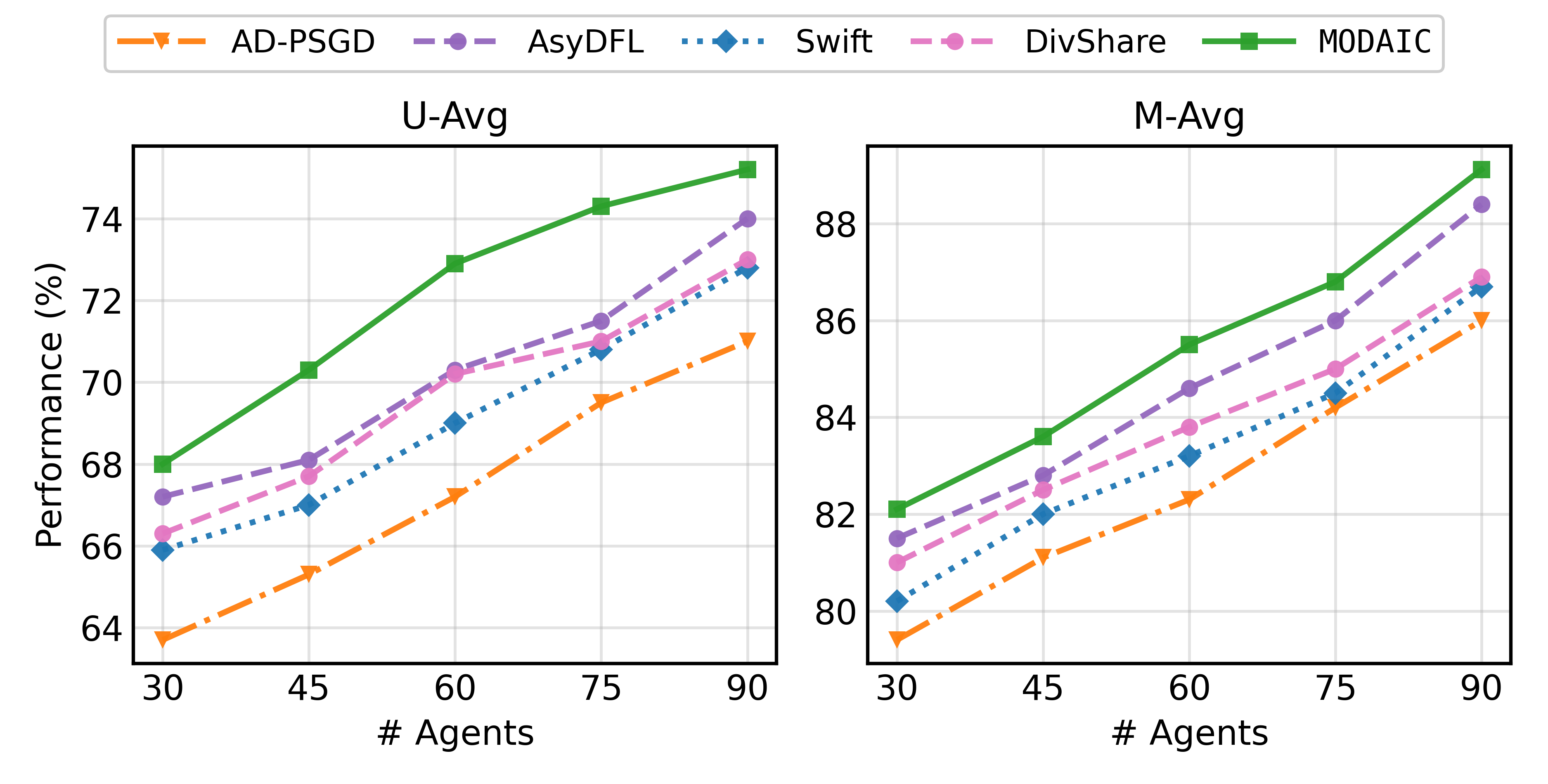}
    \vspace{-0.1in}
     \caption{\textbf{Scalability on KU-HAR.} The number of agents increases from 30 to 90. We pre-split the dataset into 90 shards so adding agents also increases available training data. Bandwidth is 1000Mbps and the multimodal fraction stays at one third.}
    \label{fig:scalability}
  \vspace{-0.15in}
\end{figure}

\noindent\textbf{Scalability.}
We scale the system from 30 to 90 agents on KU-HAR under a high-bandwidth setting (1000Mbps). To keep per-agent data size comparable as we increase the number of agents, we pre-split KU-HAR into 90 shards and assign one shard per agent.  We also keep the fraction of multimodal agents fixed at one third. Figure~\ref{fig:scalability} reports average unimodal (U-Avg) and multimodal (M-Avg) accuracy. All methods improve with more participating agents, but \MODAIC remains best at every scale: U-Avg rises from $\sim$68\% (30 agents) to $\sim$75\% (90 agents) and M-Avg from $\sim$82\% to $\sim$89\%. Notably, the gap widens at larger scales (e.g., at 90 agents, +4.3 U-Avg and +3.0 M-Avg over AD-PSGD), indicating that \MODAIC's coordination and scheduling remain effective as the network grows.

\begin{figure}[t]
    \centering
    \begin{subfigure}[t]{0.48\textwidth}
        \centering
        \includegraphics[width=0.65\linewidth]{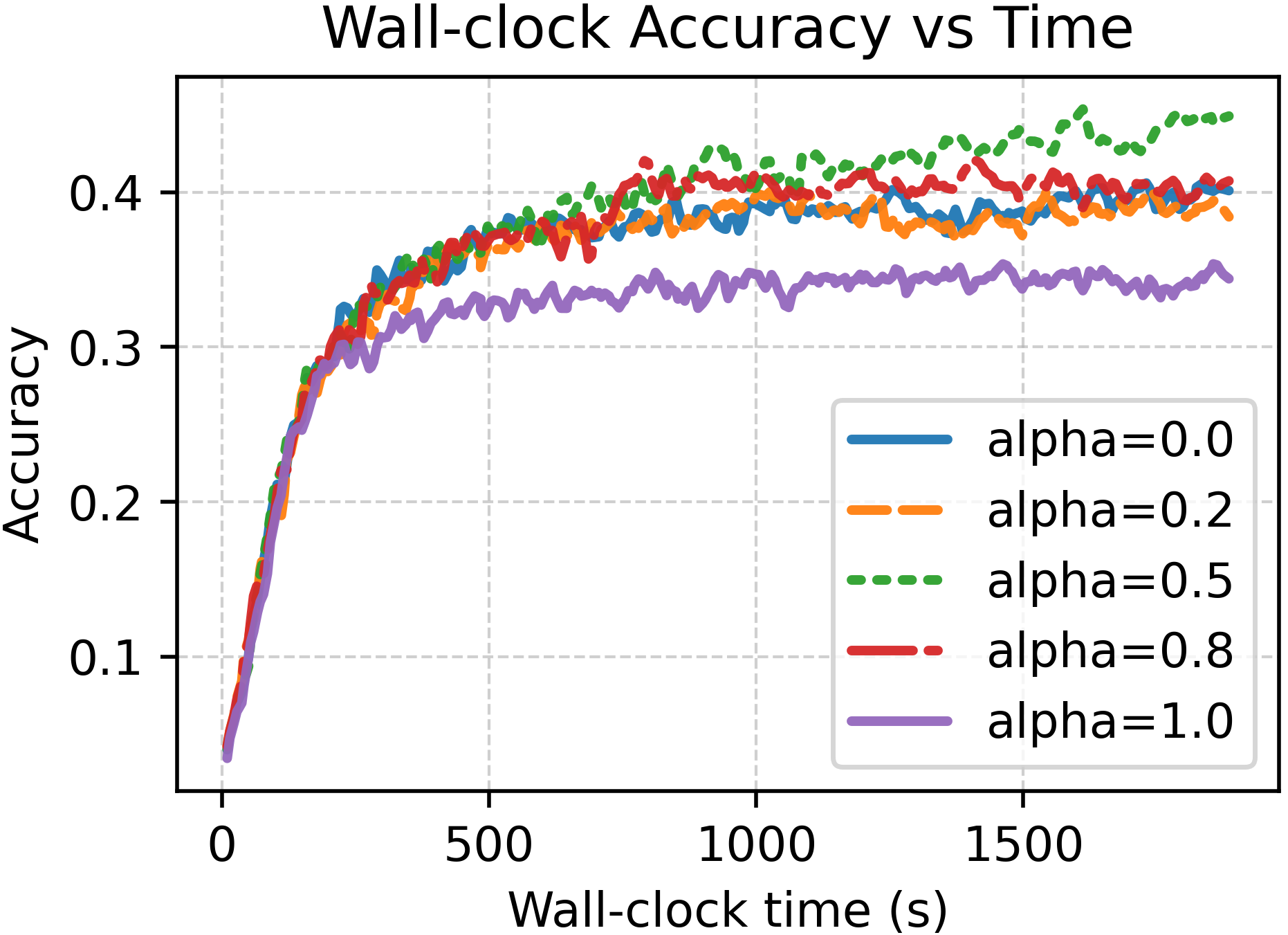}
        \caption{Different $\alpha$.}
    \end{subfigure}%
    \begin{subfigure}[t]{0.48\textwidth}
        \centering
        \includegraphics[width=0.65\linewidth]{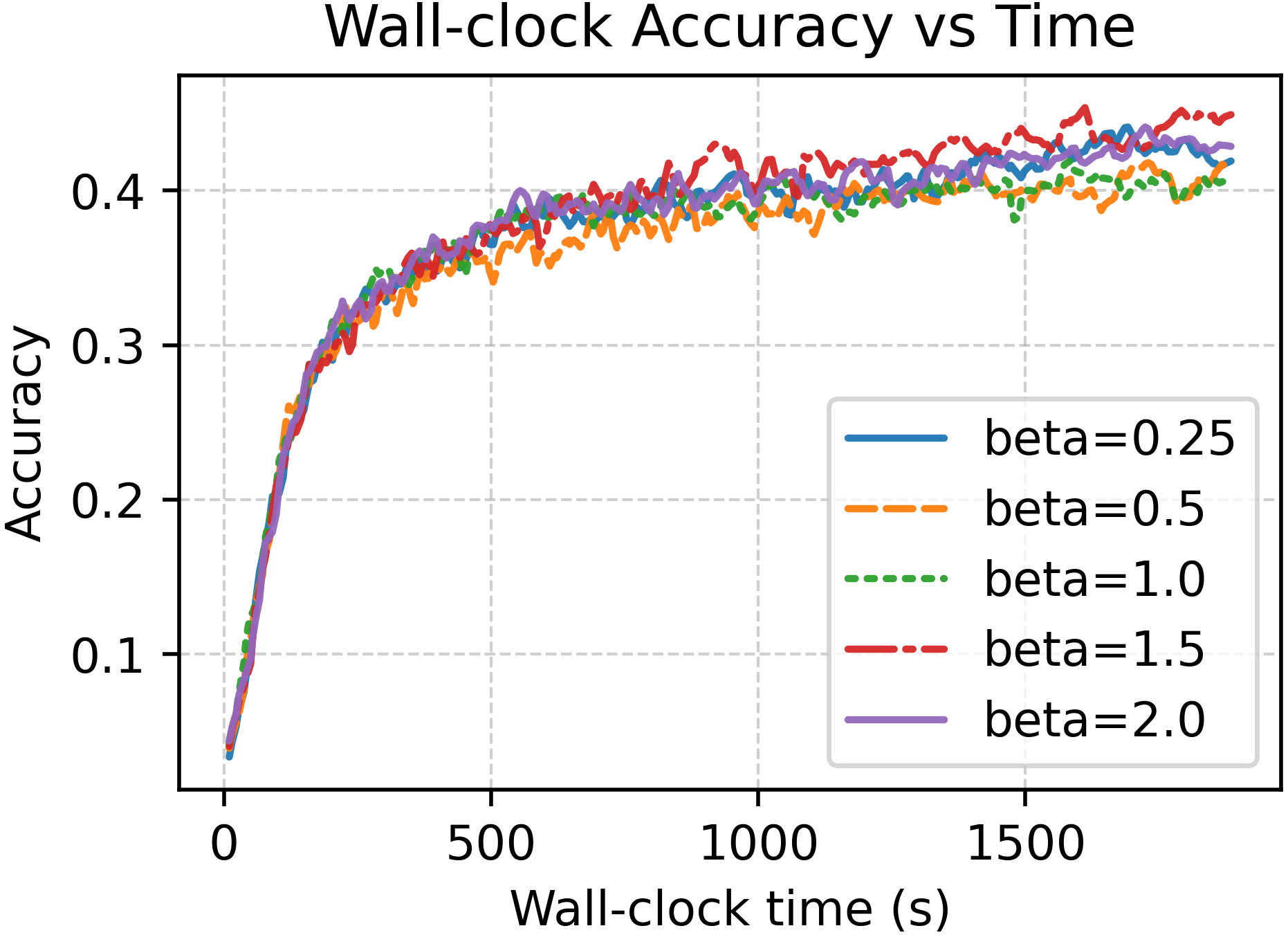}
        \caption{Different $\beta$.}
    \end{subfigure}
    \vspace{-0.1in}
    \caption{Sensitivity of consensus-aware peer selection scoring hyperparameters ($\alpha$,$\beta$) on unimodal performance (AVE).}
    \label{fig:alpha_beta_sweep}
    \vspace{-0.15in}
\end{figure}

\noindent\textbf{Staleness–fairness trade-off.}
To evaluate whether our peer selection design achieves a reasonable staleness–fairness trade-off, and how this trade-off impacts accuracy, we sweep the hyperparameters $\alpha$ and $\beta$ in our scoring function. As shown in Figure~\ref{fig:alpha_beta_sweep}, varying $\alpha$---which controls the relative weight between the fairness term (how often a peer has been selected) and the loss-similarity term---leads to a clear change in convergence behavior. When $\alpha = 0.5$, the unimodal performance is highest, suggesting that, at our scale, staleness control (via fairness) and similarity-based mixing are roughly equally important for convergence.
We also observe that $\beta = 1.5$ yields the best performance among the tested values. Recall that $\beta$ acts as a temperature parameter in the loss-similarity kernel: smaller $\beta$ makes the kernel overly sharp and restricts aggregation to only very similar peers, while larger $\beta$ flattens the kernel and makes selection almost uniform. The optimal value around $\beta=1.5$ indicates that our system benefits from a moderately smooth similarity kernel that still prefers similar peers, but does not exclude slightly less similar ones.

\begin{table*}[t]
\centering
\small
\setlength{\tabcolsep}{4pt}
\renewcommand{\arraystretch}{1.15}
\begin{tabular}{lcccc|cccc}
\toprule
\multirow{2}{*}{\textbf{Method}} &
\multicolumn{4}{c}{\textbf{Dir. $\alpha=0.1$}} &
\multicolumn{4}{c}{\textbf{Dir. $\alpha=10.0$}} \\
\cmidrule(lr){2-5}\cmidrule(lr){6-9}
& \textbf{U-Avg} & \textbf{U-Worst} & \textbf{M-Avg} & \textbf{M-Worst}
& \textbf{U-Avg} & \textbf{U-Worst} & \textbf{M-Avg} & \textbf{M-Worst} \\
\midrule
AD-PSGD   & 21.10 & 16.24 & 26.58 & 18.06  & 34.53 & 23.72 & 47.67 & 38.48 \\
AsyDFL    & \heatmaplevel{3}{30.18} & \heatmaplevel{3}{23.25} & \heatmaplevel{3}{45.34} & 31.78  & 43.12 & 37.54 & \heatmaplevel{3}{64.25} & \heatmaplevel{3}{58.14} \\
Swift     & 22.13 & 17.10 & 42.27 & 30.62  & 42.77 & 22.36 & 61.25 & 46.37 \\
DivShare  & 25.57 & 20.28 & 44.03 & \heatmaplevel{3}{33.29}  & \heatmaplevel{3}{43.83} & \heatmaplevel{3}{38.56} & 63.60 & 45.14 \\
\MODAIC &
\heatmaplevel{6}{\textbf{31.25}} & \heatmaplevel{6}{\textbf{24.17}} & \heatmaplevel{6}{\textbf{47.52}} & \heatmaplevel{6}{\textbf{44.15}} &
\heatmaplevel{6}{\textbf{50.36}} & \heatmaplevel{6}{\textbf{47.38}} & \heatmaplevel{6}{\textbf{70.14}} & \heatmaplevel{6}{\textbf{67.82}} \\
\bottomrule
\end{tabular}
\caption{\textbf{AVE ablation on data heterogeneity at 1000s under 200Mbps.}
We vary Dirichlet $\alpha$ and report U-Avg, U-Worst, M-Avg, and M-Worst across agents.}
\label{tab:ave_dir_ablation_grouped}
\vspace{-0.15in}
\end{table*}

\noindent\textbf{Data heterogeneity.} Using the default bandwidth budget, we evaluate different methods on AVE under varying data heterogeneity levels (Dirichlet parameter $\alpha$) in Table~\ref{tab:ave_dir_ablation_grouped}. When heterogeneity is mild (i.e., $\alpha=10.0$), our method consistently improves both average and worst-case performance for unimodal and multimodal metrics. Under more severe heterogeneity (smaller $\alpha$), our method still provides stronger worst-case multimodal performance, indicating improved robustness to highly skewed client data.

\subsection{Details on Cloud-based Experiments}

\textbf{Cloud setting.}
We deploy a cross-region cloud testbed on AWS using three EC2 instances located in Virginia (US), London (UK) and Singapore, each provisioned with 64 vCPUs and 256 GB RAM, and run all agents as Docker containers (CPU-only). To enable container-to-container IP connectivity across regions, we build an encrypted host overlay using WireGuard in a full-mesh configuration; each host advertises a unique Docker bridge subnet and forwards remote container subnets through the WireGuard interface, yielding stable private container IPs for end-to-end TCP communication. We orchestrate experiments from an external controller (a local server) via SSH, which launches/stops containers on each host, assigns roles (tracker vs.\ worker), and collects logs from host-mounted volumes, while we also allocate CPU shares to worker containers using cgroup quotas and apply per-container bandwidth caps with Linux traffic control (HTB/\texttt{netem}) on each container interface to emulate heterogeneous wide-area links.

\begin{figure}[t]
    \centering
    \includegraphics[width=.65\linewidth]{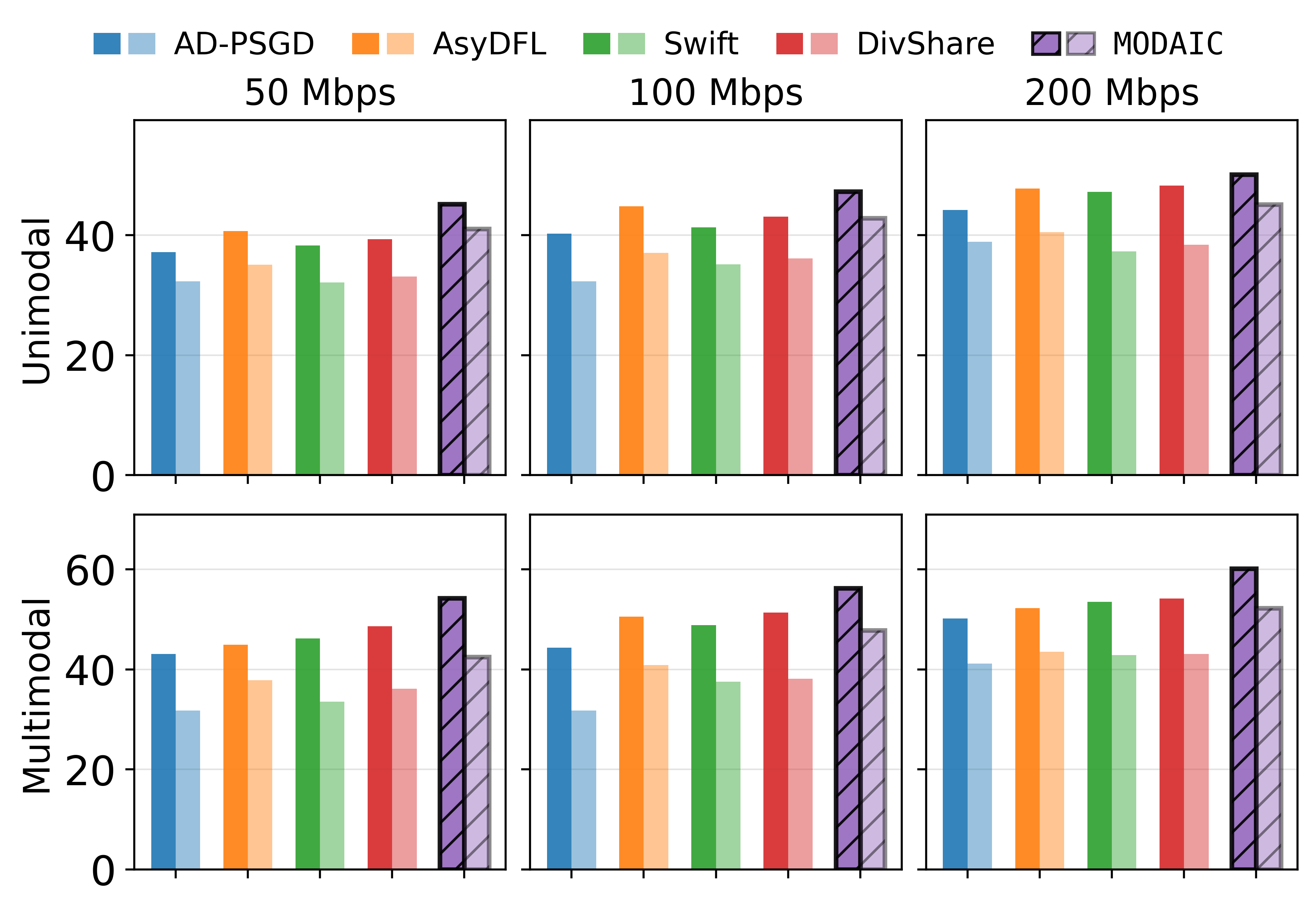}
    \vspace{-0.1in}
     \caption{\textbf{Cloud bandwidth sweep on AVE} (CPU-only, 3 AWS regions). Columns vary per-agent bandwidth cap. Rows show unimodal (top) and multimodal (bottom) accuracy. Bars show group average and worst-agent accuracy. Results are measured at 5000s.
     }
    \label{fig:bandwidth_sweep_aws}
    \vspace{-0.15in}
\end{figure}

\subsection{Additional Wide-Area Cloud Deployment}\label{sec:cloud_exp_app}

\noindent\textbf{Bandwidth sweep.}
Figure~\ref{fig:bandwidth_sweep_aws} reports unimodal and multimodal average/worst accuracy under different bandwidth budgets. Unlike our GPU-based local testbed, cloud runs are CPU-bound, so reducing bandwidth from 200Mbps to 50Mbps causes a smaller accuracy drop. The average-worst gap is more pronounced because limited CPU parallelism amplifies compute-time skew across agents. Despite this shift in the bottleneck, \MODAIC consistently achieves the best average and worst-agent accuracy for both unimodal and multimodal groups.

\section{Final Convergence Accuracy} \label{sec:convergence_app}

\begin{figure}[ht]
  \centering
  \includegraphics[width=0.65\linewidth]{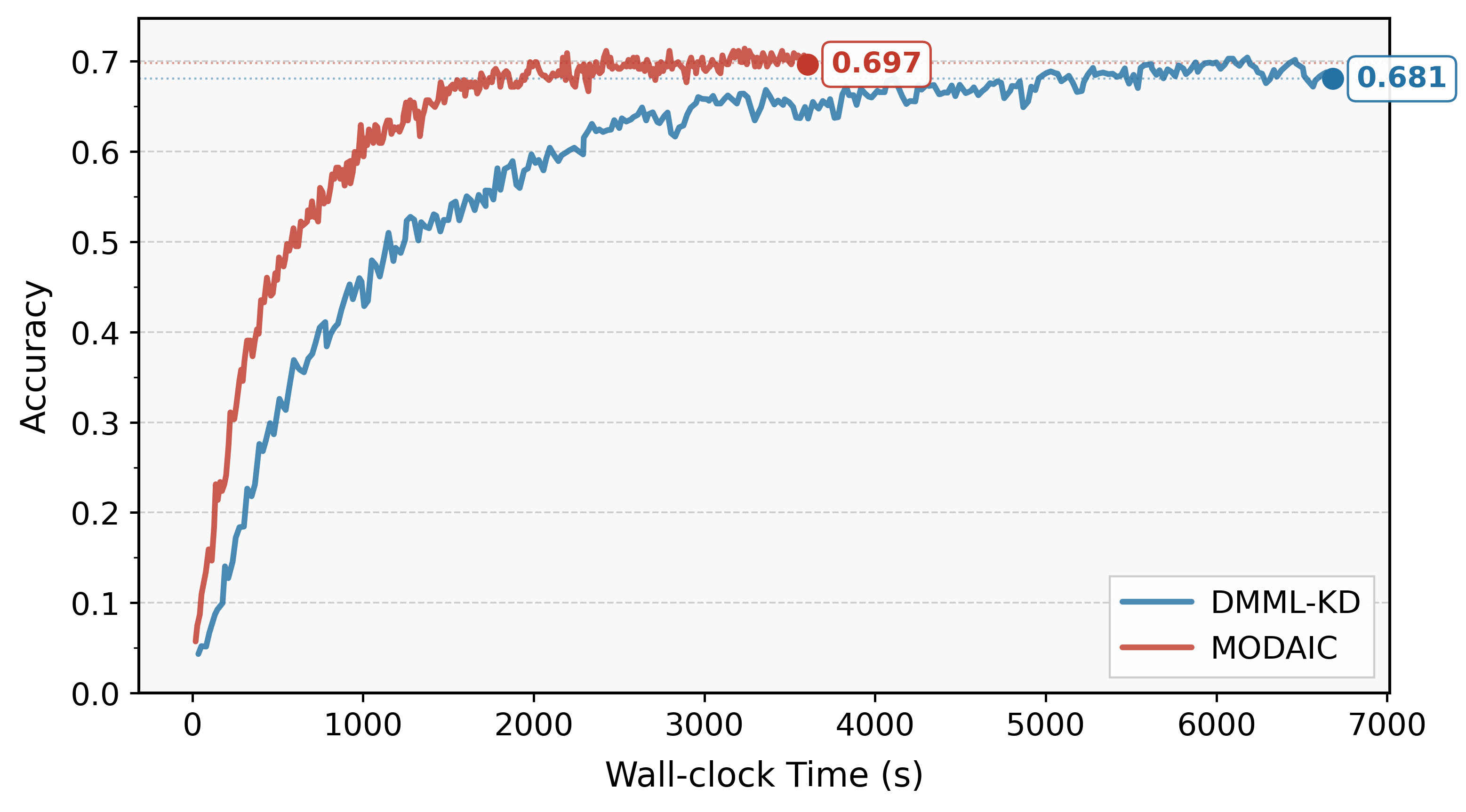}
  \caption{Wall-clock accuracy on AVE until saturation.
           \textsc{Modaic} (red) converges to
           69.8\%; \textsc{Dmml-KD} (blue)
           converges to 68.1\%.
           Dotted horizontal lines mark each method's final accuracy ceiling.}
  \label{fig:modaic_convergence}
\end{figure}

Figure~\ref{fig:modaic_convergence} plots wall-clock accuracy on the AVE benchmark
for \MODAIC and DMML-KD run until saturation.
Both curves plateau at convergence: \MODAIC reaches a final accuracy of 69.8\%
while DMML-KD reaches 68.1\%, confirming that \MODAIC's gains are not merely
early-time artifacts but persist as a higher final accuracy ceiling compared to the best synchronized baseline.

\end{document}